\documentclass[12pt]{iopart}
\expandafter\let\csname equation*\endcsname\relax
\expandafter\let\csname endequation*\endcsname\relax
\usepackage{amsfonts} 
\usepackage{amsmath}
\usepackage{iopams}
\usepackage{amssymb}
\usepackage{mathtools}
\usepackage{cancel} 
\usepackage{bm}
\usepackage[usenames, dvipsnames]{color} 
\usepackage{dcolumn}
\usepackage{epic} 
\usepackage{epsfig}
\usepackage{feynmp}
\usepackage{graphicx}
\usepackage{grffile}
\usepackage[breaklinks,colorlinks = true,linkcolor = red,urlcolor=blue,citecolor=red]{hyperref}
\usepackage{mathrsfs}
\usepackage{soul}
\usepackage{wrapfig}
\usepackage{xy} 
\usepackage{xcolor}
\usepackage{amsmath,amssymb,amsfonts,amsthm}
\usepackage[ascii]{inputenc}
\usepackage{makecell}
\allowdisplaybreaks

\newcommand{\be}{\begin{equation}}
	\newcommand{\ee}{\end{equation}}
\newcommand{\ba}{\begin{aligned}}
	\newcommand{\ea}{\end{aligned}}
\newcommand{\bea}{\begin{eqnarray}}
	\newcommand{\eea}{\end{eqnarray}}	
	
\newcommand{\beal}{\begin{align}}
\newcommand{\eal}{\end{align}}

\graphicspath{{images/}{../Figures}}

\begin{document}

\title{Truncated linear statistics in the one dimensional one-component plasma}

\author{Ana Flack}
\address{LPTMS, CNRS, Univ.  Paris-Sud,  Universit\'e Paris-Saclay,  91405 Orsay,  France}
\author{Satya N. Majumdar}
\address{LPTMS, CNRS, Univ.  Paris-Sud,  Universit\'e Paris-Saclay,  91405 Orsay,  France}
\author{Gr\'egory Schehr}
\address{Sorbonne Universit\'e, Laboratoire de Physique Th\'eorique et Hautes Energies, CNRS UMR 7589, 4 Place Jussieu, 75252 Paris Cedex 05, France}



\begin{abstract}
In this paper, we study the probability distribution of the observable $s = (1/N)\sum_{i=N-N'+1}^N x_i$, with $1 \leq N' \leq N$ and  
$x_1<x_2<\cdots< x_N$ representing the ordered positions of $N$ particles in a $1d$ one-component plasma, i.e., $N$
harmonically confined charges on a line, with pairwise repulsive $1d$ Coulomb interaction $|x_i-x_j|$. This observable represents
an example of a truncated linear statistics -- here the center of mass of the $N' = \kappa \, N$ (with $0 < \kappa \leq 1$), rightmost particles.
It interpolates between the position of the rightmost particle (in the limit $\kappa \to 0$) and the full center of mass (in the limit $\kappa \to 1$). We show that, for large
$N$, $s$ fluctuates around its mean $\langle s \rangle$ and the typical fluctuations are Gaussian, of width $O(N^{-3/2})$. The atypical large fluctuations of $s$, for fixed $\kappa$, are instead described by a large deviation form ${\cal P}_{N, \kappa}(s)\simeq \exp{\left[-N^3 \phi_\kappa(s)\right]}$, where
the rate function $\phi_\kappa(s)$ is computed analytically. We show that $\phi_{\kappa}(s)$ takes different functional forms in five distinct
regions in the $(\kappa,s)$ plane separated by phase boundaries, thus leading to a rich phase diagram in the $(\kappa,s)$ plane. Across
all the phase boundaries the rate function $\phi(\kappa,s)$ undergoes a third-order phase transition. This rate function
is also evaluated numerically using a sophisticated importance sampling method, and we find a perfect agreement with our analytical predictions.  
\end{abstract}
\date{\today}
\maketitle

\maketitle
\section{Introduction}\label{Introduction}

A plasma in one-dimension consists of an equal number of opposite charges, interacting via the $1d$-Coulomb interaction.
Denoting by $x_i$'s and $y_i$'s the positions of the two species of opposite charges, with $i=1,2, \cdots, N$, the energy
of the plasma reads~\cite{Lenard,Prager,Baxter,Choquard}
\bea \label{def_Eplasma}
E[\{x_i\}, \{y_i\}] = -B \sum_{i \neq j} |x_i-x_j| - B \sum_{i \neq j} |y_i-y_j| + B\, \sum_{i \neq j} |x_i-y_j| \;,
\eea
where the charges of each species repel each other via the linear Coulomb potential in $1d$, while opposite charges
attract each other via the Coulomb attraction. The coupling $B>0$ just denotes the strength of the interaction. Instead of
treating both species of charges microscopically, an useful approximation, valid in the large $N$ limit, is to treat only one of them (say the $x_i$'s) microscopically,
while treating the other species (the $y_i$'s) as a uniform background density $\rho_0$ of opposite charges. This background density is
supported over a finite symmetric interval $[-L,L]$, such 
that $\rho_0\times 2L = N$, maintaining the overall
charge neutrality.  The coupling term $B\, \sum_{i \neq j} |x_i-y_j|$ in Eq. (\ref{def_Eplasma}) can then be approximated as~\cite{Lenard,Prager,Baxter,Choquard}
\bea \label{coupling}
B\, \sum_{i \neq j} |x_i-y_j| \simeq B\, N\, \rho_0 \sum_{i=1}^N \int_{-L}^L |x_i-y| \, dy = B\, N\, \rho_0 \sum_{i=1}^N \left( L^2 + x_i^2\right)  \;.
\eea
Thus the uniform background of negative charges gives rise to an effective harmonic confining potential for the positive charges. Dropping all the constant terms, the
microscopic energy of the $x_i$'s can then be written as
\bea \label{def_Ex}
E[\{ x_i\}] =  A \sum_{i=1}^N x_i^2 -B \sum_{i \neq j} |x_i-x_j|  \;,
\eea
where $A$ and $B$ are positive constants. Thus this effective model for the positive charges corresponds to a $1d$-Coulomb gas in the presence of 
a harmonic potential. This is what is referred to as the $1d$ one-component plasma ($1d$OCP), also known as the jellium model~\cite{Lenard,Prager,Baxter,Choquard}. The harmonic term tries to push the charges close to the origin, while
the repulsive interaction tries to spread them apart. The competition between these two terms leads to interesting collective properties of these charges~\cite{Choquard,Dean,Tellez,Dhar2017,Dhar2018,cunden2018universality}.

Given this energy function in Eq. (\ref{def_Ex}), we are interested in the equilibrium properties of the system, where the probability to find a specific
configuration $\{x_i\}$ is given by the Boltzmann distribution
\bea \label{boltz}
{\cal P}(\{ x_i\}) = \frac{e^{-\beta E[\{x_i\}]}}{Z_N} \;,
\eea
where $\beta$ is the inverse temperature and $Z_N$ is the partition function, that normalizes this probability distribution 
\bea \label{ZN}
Z_N = \int dx_1  dx_2 \cdots  dx_N \, e^{-\beta E[\{x_i\}]}  \;.
\eea
We are interested in the large $N$ limit where this multiple integral in Eq. (\ref{ZN}) is expected to be dominated by the maximum
of the integrand, i.e., by the minimal energy configuration (the ground-state). To bring out the explicit $N$-dependence of the energy, it
is useful to estimate how the two terms in the energy in Eq. (\ref{def_Ex}) scale with $N$. We start with the first term. Suppose that $x_i \simeq L_N \tilde x_i$
where $\tilde x_i = O(1)$. Then the first term $E_1 =  A \sum_{i=1}^N x_i^2$ scales for large $N$ as $E_1 \simeq A\,L_N^2\, N$ where the factor $N$ comes from the fact that there are $N$ terms
of order $O(1)$ in the summation in $E_1$. Similarly, the second term $E_2 = B \sum_{i \neq j} |x_i-x_j|$ scales as $E_2 \simeq B\, L_N\, N^2$ where the factor $N^2$ comes from the fact that there
are $N(N-1)$ pairwise terms of order $O(1)$ in the double sum in $E_2$. Since we want both terms of the energy $E_1$ and $E_2$ to compete with each other, they should be of the same order. This leads to
\bea \label{LN_1}
A \, L_N^2\, N \simeq B \, L_N \, N^2 \Longrightarrow L_N =  O(N) \;.
\eea 
Therefore, the scaled dimensionless energy $\beta E[\{x_i\}]$ in Eq. (\ref{ZN}) can be expressed as
\bea \label{scaledE}
\beta E[\{x_i\}] = \frac{N^2}{2} \sum_{i=1}^N \tilde x_i^2 - \alpha \,N\, \sum_{i\neq j} |\tilde x_i - \tilde x_j| \;,
\eea
where $\tilde x_i = O(1)$. We have chosen $A$ and $B$ in Eq. (\ref{def_Ex}) appropriately and the positive $\alpha = O(1)$ denotes the effective interaction strength. For convenience, we
will henceforth drop the notation $\tilde x_i$ and replace it with $x_i$
\bea \label{scaledE_2}
\beta E[\{x_i\}] = \frac{N^2}{2} \sum_{i=1}^N x_i^2 - \alpha \,N\, \sum_{i\neq j} |x_i - x_j| \;.
\eea

Let us first find out the minimal energy configuration in the large $N$ limit. For this, it is convenient to re-write the partition function in terms of the ordered positions 
$x_1<x_2<\cdots< x_N$
\bea 
Z_N &=& N !  \int dx_1  \,dx_2 \cdots  \, dx_N \, e^{-\beta E[\{x_i\}]} \prod_{j=2}^N \theta(x_j - x_{j-1}) \label{ZN_ord} \\
&=& N !  \int_< dx_1  \, dx_2 \cdots  dx_N  \; e^{-\beta E[\{x_i\}]}  \;, \label{ZN_ord2}
\eea
where the factor $N!$ comes from the fact that the energy function is symmetric under the permutations of the $x_i$'s and the subscript '$<$' in Eq. (\ref{ZN_ord2}) is a short-hand notation for the integration over the ordered sector $x_1<x_2< \cdots x_N$. The advantage of the ordering is that we can replace
$|x_i - x_j| = (x_i-x_j)$ for $i>j$. In that case, it is easy to show that the energy function~(\ref{scaledE_2}) reads
\be
 \beta E(\{x_i\})=\frac{N^2}{2}\sum_{i=1}^{N}\left(x_i-\left[ \frac{2\alpha}{N}(2i-N-1)\right]\right)^2-2\alpha^2\sum_{i=1}^{N}(2i-N-1)^2 \;.
 \label{eq:Ham2}
 \ee 
 Minimizing this energy function trivially, one finds the ground state configuration where the charge positions are given by \cite{Lenard,Prager,Baxter}
 \bea \label{E_GS}
 x_i = x_i^* = \frac{2\alpha}{N}(2i-N-1) \quad, \quad {\rm for} \quad i = 1, 2, \cdots N \;.
 \eea
 The corresponding energy from Eq. \eqref{eq:Ham2} is $\beta E(\{x_i^*\}) = -2\alpha^2\sum_{i=1}^{N}(2i-N-1)^2 = - (2\alpha^2)(N^3-N)/3 \simeq - (2\alpha^2)/3\, N^3$ for $N \gg 1$. Evaluating the partition function $Z_N$ in Eq. \eqref{ZN_ord2} by a saddle-point (using the minimum energy configuration) gives the leading order behavior
 \bea \label{ZNstar}
 Z_N = e^{\frac{2 \alpha^2}{3}\, N^3 + O(N)} \;.
 \eea

\begin{figure}
\centering
\includegraphics[width = 0.7\linewidth]{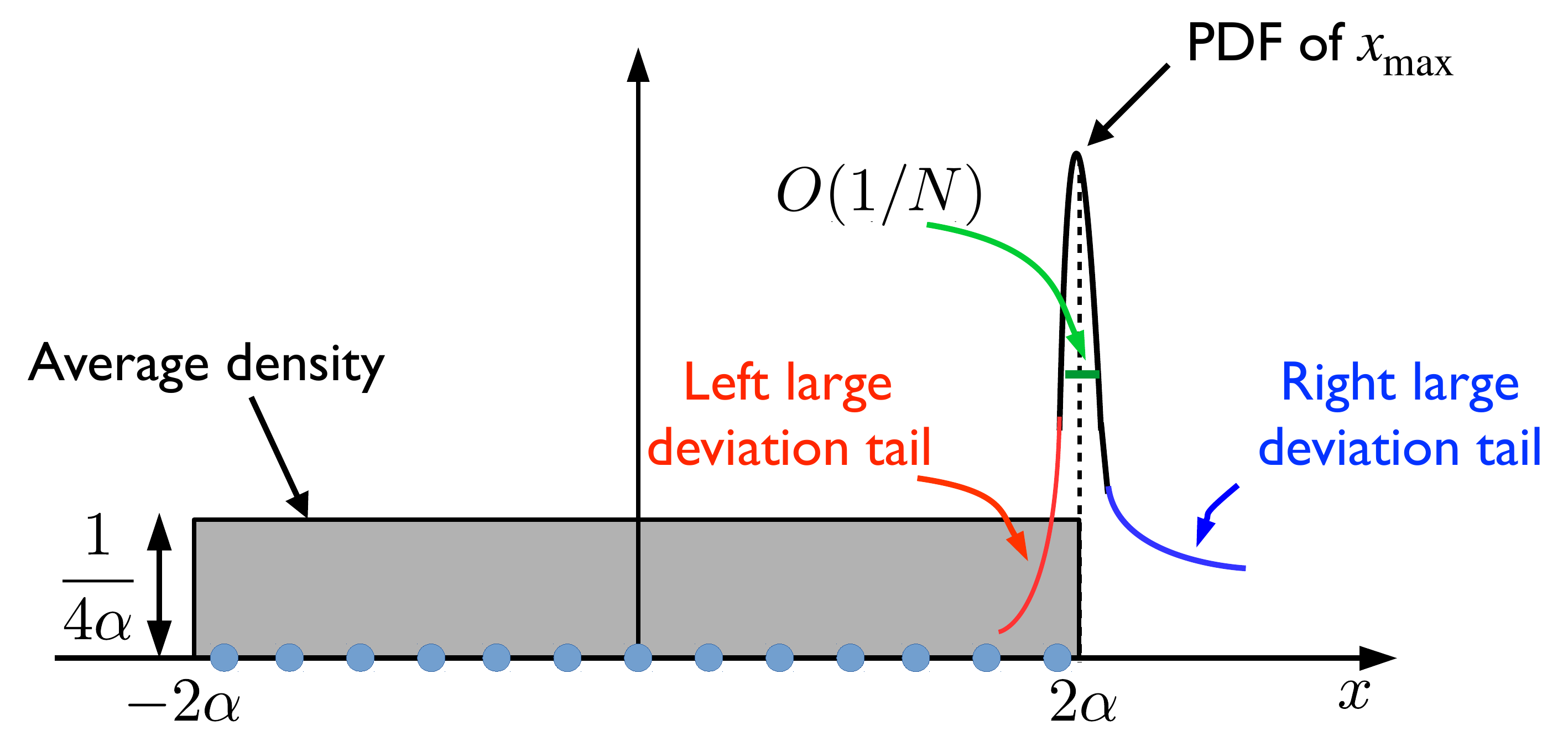}
\caption{The shaded region shows the flat average density $\rho(x) = 1/(4\alpha)$ inside the support $[-2\alpha, + 2\alpha]$. The probability distribution function (PDF) of the position $x_{\max}$ of the rightmost particle is shown schematically. This PDF is highly peaked around the mean position $2\alpha$ of $x_{\max}$. The typical fluctuations of $x_{\max}$ around its mean $2 \alpha$ is of order $O(1/N)$, as indicated in the figure. The atypical large fluctuations of $x_{\max}$, of order $O(1)$ around the mean $2 \alpha$, on its left and right, are described respectively by the left and the right large deviation functions, see Eq. (\ref{Q_largeN}).} \label{Fig_xmax}
\end{figure} 
Going back to the configuration with minimum energy in Eq. \eqref{E_GS}, we see that the charges are equally spaced, with interval $4 \alpha/N$. The rightmost charge is located at $x_N^* = 2\alpha(1-1/N)$ and the leftmost
at $x_1^* = - 2\alpha(1-1/N)$. Thus the macroscopic charge density, in the large $N$ limit, becomes flat on the support $[-2\alpha, +2\alpha]$ and is strictly zero outside (see Fig. \ref{Fig_xmax})
\begin{align}
 \rho(x)=\left\{
 \begin{array}{ll}
 	\frac{1}{4\alpha}, \quad -2\alpha<x<2\alpha \;,\\
 	0, \quad \quad \textnormal{elsewhere} \;.
 \end{array}
 \right.
 \label{eq:eq_density}
\end{align}
For finite but large $N$, the position of the rightmost particle $x_{\max} = x_N$ fluctuates around its mean value $2 \alpha$ and the scale of these typical fluctuations 
are of order $O(1/N)$. On this scale, the PDF is described by a scaling form 
\begin{align}\label{pdf_xmax}
{\rm Prob.}(x_{\max} = w, N) \simeq N \, f_{\alpha}(N(w-2\alpha)) \;,
\end{align}
where the scaling function $f_{\alpha}(x)$ was computed recently \cite{Dhar2017, Dhar2018} and it has asymmetric tails
\bea \label{asymptFa}
f_\alpha(x) \simeq 
\begin{cases}
&\exp{[-|x|^3/(24 \alpha) + O(x^2)]} \quad\,, \quad {\rm as} \quad \quad x \to - \infty\\
&\exp{[-x^2/2+O(x)]} \quad \quad \quad \quad, \quad  {\rm as} \quad \quad  x \to + \infty \;.
\end{cases}
\eea
However, atypical large fluctuations where $x_{\max} - 2\alpha = O(1)$ on the two sides of the mean are not described by $f_\alpha(x)$ but
instead are described by large deviation forms. Indeed, the full function of the PDF can be summarised as follows (see Fig. \ref{Fig_xmax})
\begin{align} \label{Q_largeN}
{\rm Prob.}(x_{\max} = w, N)  \simeq
\begin{cases}
&e^{-N^3 \, \Phi_-(w)+ O(N^2)} \;, \; \quad\quad 0<2\alpha - w = O(1) \\
&N\,f_\alpha(N(w-2\alpha))\;,\, \quad \; |w - 2 \alpha| = O(1/N) \\
&e^{-N^2 \Phi_+(w)+O(N)}\;, \quad \quad \;\; 0<w-2\alpha = O(1) \;,
\end{cases}
\end{align}
where the left and the right large deviation functions $\Phi_{-}(w)$ and $\Phi_+(w)$ 
were also computed explicitly~\cite{Dhar2017, Dhar2018}. The left rate function is given by~\cite{Dhar2017, Dhar2018}
\begin{eqnarray}\label{left_rf}
\Phi_{-}(w) = 
\begin{cases}
& \dfrac{(2\alpha-w)^3}{24\alpha}\;,\;\quad -2\alpha\leq w \leq 2\alpha \\
& \\
& \dfrac{w^2}{2} + \dfrac{2}{3} \alpha^2 \;, \;\quad\; w \leq -2\alpha \;.
\end{cases}
\end{eqnarray}
while the right one is simply~\cite{Dhar2017, Dhar2018}
\begin{eqnarray}
\Phi_+(w)& =& \frac{(w-2\alpha)^2}{2}\;,\;\quad w \geq 2\alpha \;. \label{right_rf} 
\end{eqnarray}
Thus, as a function of $w$ and on a scale $w = O(1)$, there are two phase transitions in the large deviation form, respectively at $w=2 \alpha$ and
$w = - 2 \alpha$. We will come back to the details of these phase transitions at a later stage.

In this paper, we are interested not just on the position rightmost particle, but on the sum of the positions of the $N'$ rightmost particles where we set $N' = \kappa \,N$, so that $0<\kappa \leq 1$. We denote this quantity (scaled by $1/N$ to keep its typical value to order $O(1)$) by
\be 
s=\frac{1}{N}\sum_{i=N-N'+1}^{N} x_i = \frac{1}{N}\sum_{i=N(1-\kappa)+1}^{N} x_i \;,
\label{eq:TLS}
\ee
where the $x_i$'s denote the {\it ordered} positions of the particles with $x_1<x_2<\cdots < x_N$. The observable $s$ in Eq. \eqref{eq:TLS} is 
clearly a random variable. The average value $\langle s \rangle$ of this random variable is easy to compute, from the equilibrium configuration (\ref{E_GS}). It reads
\bea \label{s_av}
\langle s \rangle = \frac{1}{N} \sum_{i=N-N'+1}^N \langle x_i \rangle \simeq    \frac{2\alpha}{N^2} \sum_{i=N-N'+1}^N (2i-N-1) = 2 \alpha \kappa(1-\kappa) \;,
\eea
where we have used that $\langle x_i \rangle \simeq x_i^*$ in the large $N$ limit and $N' = \kappa\,N$. However, in this paper, we are interested in the full probability distribution 
defined as
\be
\mathcal{P}_{N,\kappa}(s)= N!\, \int_< dx_1 \cdots dx_N \; \mathcal{P}(\{x_i\})\;\delta\left ( s-\frac{1}{N}\sum_{j=N(1-\kappa)+1}^{N}x_j\right) \;,
\label{eq:pd_s}
\ee
where $\mathcal{P}(\{x_i\})$ is given in Eq. (\ref{boltz}) with the energy function in Eq. (\ref{eq:Ham2}). The knowledge of the equilibrium density, while enough to determine the first moment $\langle s \rangle$, is not adequate to compute the
full distribution of $s$. 

Before proceeding to compute the full distribution of $s$, let us explain why we are interested in this observable $s$. 
In fact, for $N'=1$, the observable $s$ coincides with $x_{\max}/N$. The distribution of $x_{\max}$ is highly non trivial, as discussed above. 
In the opposite limit, when $N'=N$, then $s = \bar{x}$ is a full linear statistics of the $x_i$'s, namely in this case it corresponds to the center of mass $\bar{x}$. In that case, the distribution is a pure Gaussian with mean zero and variance $1/N^3$ for all $s$ and $N$ (for a simple proof see \ref{sec:appcofm}). It is then natural to ask how the distribution of $s$ changes as $N'$ varies between $1$ and $N$, interpolating between Eq. (\ref{Q_largeN}) and a pure Gaussian form. In the large $N$ limit, this distribution $\mathcal{P}_{N,\kappa}(s)$ is parameterised by $0<\kappa \leq 1$.  

Indeed, this observable $s$ is a particular case of the 
so-called ``truncated linear statistics'' (TLS) $\sum_{i=N-N'+1}^N f(x_i)$ (where $f(x)$ can be an arbitrary function), introduced in Ref. \cite{Grabsch, Grabsch_II} in the context of the Wishart-Laguerre ensemble of random matrix theory. The TLS can also be viewed as a special case of the so called ``thinned random matrix ensembles'' \cite{Bohigas06,charlier16,duits16}). The
Wishart-Laguerre ensemble corresponds to a model of repulsive charges where the particles are confined on the positive semi-axis and subjected to an external confining potential and a pairwise repulsion of the form $\log |x_i-x_j|$, as in the Dyson's log-gas~\cite{Dyson1962,Mehtabook,Forrester}. In this case, the $x_i$'s can be interpreted as the real eigenvalues of a matrix $X^\dagger\,X$ where $X$ is, in general, a rectangular $M \times N$  random matrix with Gaussian entries. In this case, the probability distribution $\mathcal{P}_{N,\kappa}(s)$ was studied in the large $N$ limit, using a Coulomb gas formalism, and interesting phases and transitions between them were found in the $(\kappa, s)$ plane~\cite{Grabsch, Grabsch_II}. It is interesting that TLS also appears in the computation of the ground-state energy of noninteracting trapped fermions in a random potential, both in the ``random energy model'', where the energy levels are independent \cite{Schawe2018} as well as for a class of correlated energy levels \cite{Kraj2019}.

In this paper, our goal is to investigate whether similar phase transitions in the $(\kappa,s)$ place occur in the jellium model, where the interaction between a pair of charges is linearly repulsive, as opposed to the logarithmic pairwise interaction in the Wishart-Laguerre ensemble discussed above. Our exact calculations for the jellium model in the large $N$ limit indeed display a very rich phase diagram in the $(\kappa, s)$ plane and also demonstrate how the two limits $\kappa \to 0$ (i.e., the distribution of $x_{\max}$) and $\kappa \to 1$ (i.e., the distribution of the center of mass) get connected. 

\begin{figure}[t]
	\centering
	\includegraphics[width=0.8\linewidth]{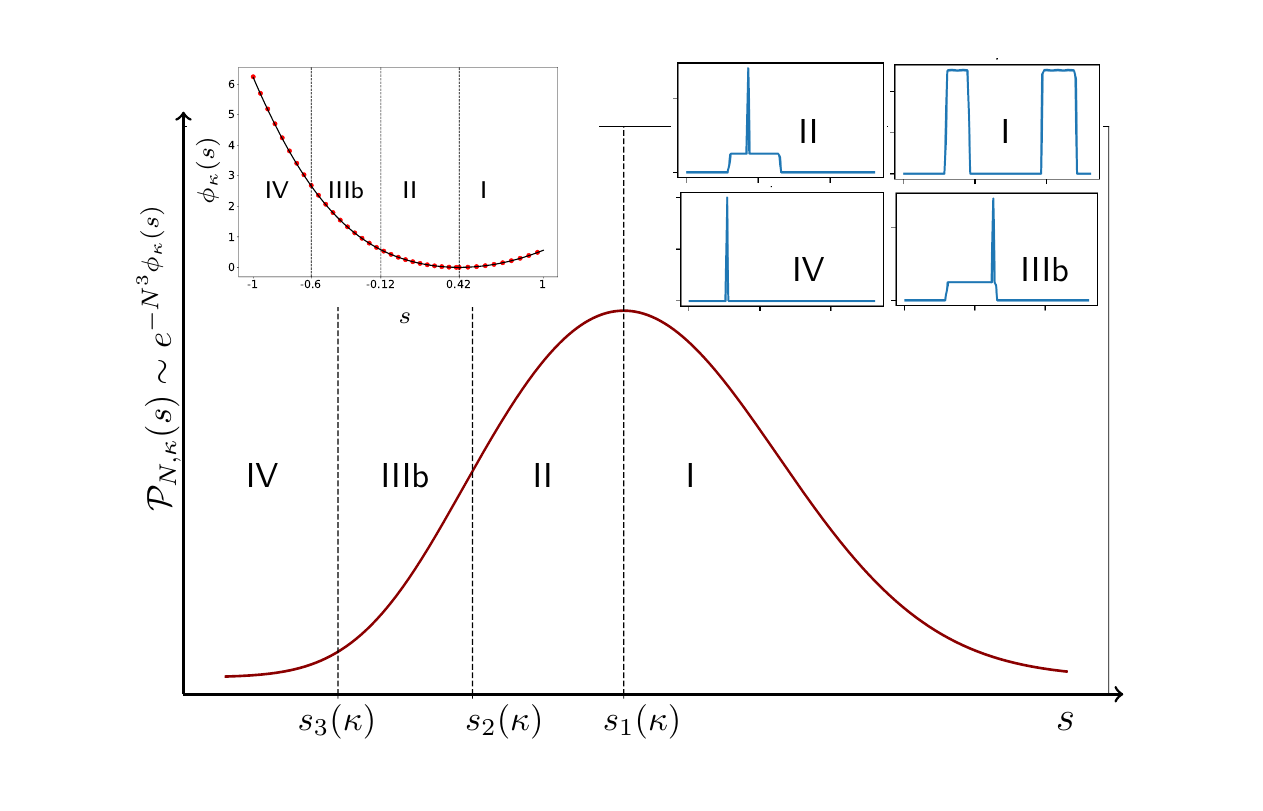}
	\caption{In the center we see a schematic depiction of the probability distribution of the truncated linear statistics $\mathcal{P}_{N,\kappa}(s)$. Four different regimes that appear for $0<\kappa \leq 0.5$ (see the text) are separated by dotted vertical lines (for $0.5\leq \kappa \leq 1$ the phase $IIIb$ is replaced by phase $IIIa$). In the upper right panel the associated saddle-point densities are plotted. In the upper left panel the large deviation function $\phi_\kappa(s)$ is plotted vs. $s$, together with the results form numerical simulation (red dots).}
	\label{fig:distr}
\end{figure}
It is useful to briefly summarise our main results. We first show that the TLS, denoted by $s$, fluctuates around its mean value $\langle s \rangle \simeq 2 \alpha \kappa(1-\kappa)$ and the typical scale of the fluctuations, for large $N$, is of order $O(N^{-3/2})$. In fact, the distribution $\mathcal{P}_{N, \kappa}(s)$, for $s - \langle s \rangle = O(N^{-3/2})$ is a pure Gaussian. This means that, on this scale, the distribution of the TLS takes the scaling form
\be \label{typ_Gaussian}
\mathcal{P}_{N, \kappa}(s) \simeq N^{3/2} f_G\left( (s-\langle s \rangle)N^{3/2}\right) 
\ee
where the scaling function $f_G(z)$ is given by
\be \label{Gaussian_form}
f_G(z) = \frac{1}{\sqrt{2 \pi \kappa}}\, e^{- \frac{z^2}{2 \kappa}} \;.
\ee
This means that the variance of $s$, for large $N$, is given by
\be \label{var_intro}
{\rm Var}(s) = \langle s^2 \rangle - \langle s \rangle^2 \simeq \frac{\kappa}{N^3} \;.
\ee
Note that, remarkably, the leading behavior of the variance, for large $N$, is independent of the interaction strength $\alpha$. 

\begin{figure}[t]
\centering
\includegraphics[width=0.6\textwidth]{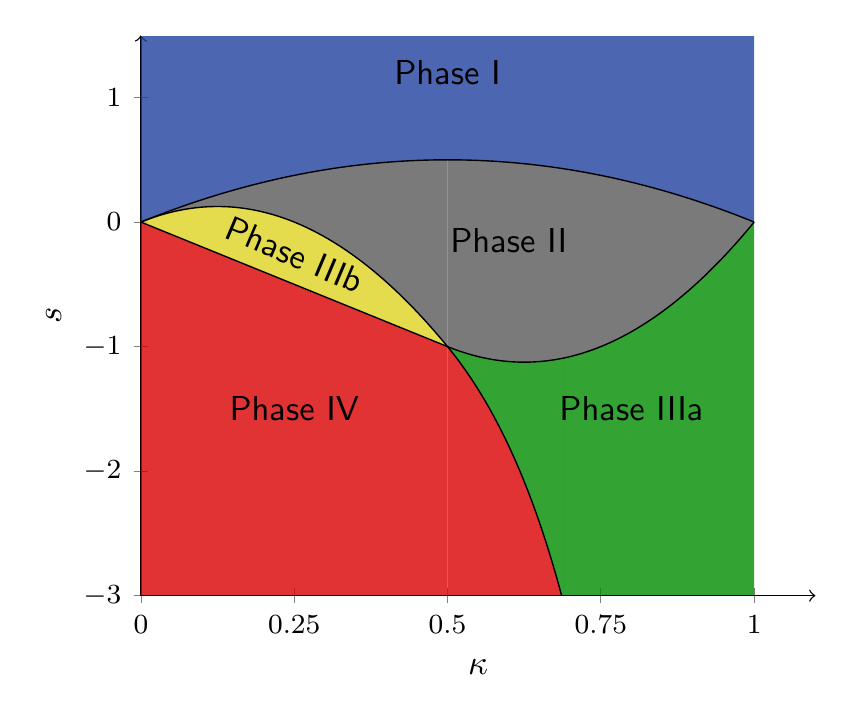}
\caption{The phase diagram in the $(\kappa,s )$ plane for $\alpha=1$. The different phases have different functional forms for the large deviation function $\phi_\kappa(s)$. For $0<\kappa\leq 1/2$ there four different phases $I, II, IIIb$ and $IV$, separated respectively by the lines $s_1(\kappa), s_2(\kappa)$ and $s_3(\kappa)$ given in Eqs. (\ref{s1})-(\ref{s3}) represented by black solid lines. Similarly, for $1/2 \leq \kappa \leq 1$, there are also four phases $I, II, IIIa$ and $IV$, separated respectively by the phase boundaries $\bar{s}_1(\kappa), \bar{s}_2(\kappa)$ and $\bar{s}_3(\kappa)$ given in Eqs. (\ref{sb1})-(\ref{sb3}) represented by black solid lines.}\label{fig:phase_space2}
\end{figure}

We next demonstrate that the atypically large fluctuations of $s$ around its mean are {\it not} described by this Gaussian form (\ref{Gaussian_form}) but rather by a large deviation form. Using a saddle-point method, valid for large $N$, 
we show that the full probability distribution of atypically large fluctuations of $s$ of order $O(1)$ around its mean admits a large deviation form
\be 
\mathcal{P}_{N, \kappa}(s)\simeq e^{-N^3\, \phi_{\kappa}(s)} \;,
\label{eq:LDF1}
\ee 
where we compute analytically the rate function $\phi_{\kappa}(s)$ as a function of $s$, for all $0<\kappa \leq 1$ -- see Eqs. \eqref{phi_k_lehalf} and \eqref{phi_k_gehalf} -- and find that $\phi_\kappa(s)$ displays a very rich behaviour as a function of $s$ for a fixed $\kappa$. This distribution $\mathcal{P}_{N, \kappa}(s)$ is plotted schematically in
Fig.~\ref{fig:distr} for $\kappa<1/2$ where we see four different regimes of $s$ (separated by vertical dotted lines). As $s$ increases and crosses the values of $s$ corresponding to 
these vertical lines (marked in the figure), the rate function $\phi_{\kappa}(s)$ exhibits a nonanalytic behavior. While the rate function and its first two derivatives are continuous across each vertical line, the third derivative is discontinuous, indicating a third-order phase transition at those critical values of $s$. This behavior is summarised in the phase diagram in the $(\kappa, s)$-plane in Fig. \ref{fig:phase_space2}, where we see a very rich phase diagram consisting of five different phases, $I, II, IIIa, IIIb$ and $IV$. In each phase, the configurations of particles that contribute dominantly to $\mathcal{P}_{N, \kappa}(s)$, for large $N$, have densities of different shapes, as shown schematically in the right top inset of Fig. \ref{fig:distr}. In addition, from our general large deviation results for ${\cal P}_{N, \kappa}(s)$, valid for all $0 < \kappa \leq 1$, we show how to recover the distribution of $x_{\max}$ and the center of mass in the two limiting cases, respectively $\kappa \to 0$ and $\kappa \to 1$.

In order to verify our analytical predictions for the rate function $\phi_{\kappa}(s)$ in Eq. (\ref{eq:LDF1}), it would be useful to compute it numerically. However, a numerical computation of $\phi_{\kappa}(s)$ is highly nontrivial, as it corresponds to probabilities that are extremely tiny $\simeq e^{-N^3}$. A normal Monte-Carlo simulation will never capture this tail. Here we employ an importance sampling method \cite{Nadal2010,Nadal2011,Hartmann2011, Hartmann2018, Banerjee2020,Mori2021} to compute this rate function extremely accurately for different values of $\kappa$. In the top left inset of Fig. \ref{fig:distr}, we have compared the numerically obtained $\phi_\kappa(s)$ (red dots) with our analytical expression (black line), for $\kappa = 0.3$. They are essentially indistinguishable. 

The rest of the paper is organised as follows. In Section \ref{sec:Laplace}, we set up the method to compute the distribution ${\cal P}_{N,\kappa}(s)$, both when the fluctuation $s-\langle s \rangle$ around its mean is ''typical'' or anomalously large. It turns out to be easier to first derive the Laplace transform, or more appropriately the cumulant generating function, (with respect to $s$) of ${\cal P}_{N,\kappa}(s)$. In Section \ref{sec:deriv}, we present the computation of the large $N$ behavior of this Laplace transform and show that it leads to a rich phase diagram shown in Fig. \ref{fig:phase_space1} in the $(\kappa, \mu)$ plane, where $\mu$ is the Laplace variable conjugate to $s$. This is analogous to studying the problem in the grand-canonical ensemble. 
In Section \ref{sec:phdiag}, we show how the results obtained in the $(\kappa, \mu)$ plane can be translated to 
the phase diagram in the $(\kappa,s)$ plane, as shown in Fig. \ref{fig:phase_space2}. In Section \ref{sec_large_dev}, we compute explicitly the large deviation function $\phi_\kappa(s)$ associated with the distribution ${\cal P}_{N,\kappa}(s)$ and show that it undergoes third-order phase transitions across the phase boundaries in the $(\kappa,s)$ plane. We then present the details of our numerical simulations in Section \ref{sec:MC}. Finally, we conclude in Section \ref{sec:conclu}. Some details of the computations are relegated to three appendices.

\section{The general setup to compute the distribution of the TLS} \label{sec:Laplace}

We start by substituting Eqs. (\ref{boltz}) and (\ref{scaledE_2}) in Eq. (\ref{eq:pd_s}) which then reads
\begin{align} \label{Pnks}
{\cal P}_{N,\kappa}(s) = \frac{N!}{Z_N}\int_< dx_1 \cdots dx_N e^{-\left[\frac{N^2}{2}\sum_{i=1}^N x_i^2 - \alpha \, N\,\sum_{i\neq i} |x_i-x_j| \right]} \, \delta\left(s - \frac{1}{N} \sum_{i=N(1-\kappa)+1}^N x_i \right) \;.
\end{align}
We now define the cumulant generating function (analogue of the Laplace or the Fourier transform) 
\begin{align}\label{LTPnks}
\hat{\mathcal{P}}_{N, \kappa}(\tilde \mu) = \int_{-\infty}^\infty {\cal P}_{N,\kappa}(s) \, e^{- \tilde \mu \,s} \, ds \;.
\end{align}
Note that here the variable $s$ can be both positive and negative and in this sense, this is more like a cumulant generating function than strictly a Laplace transform. Taking Laplace transform of Eq. (\ref{Pnks}) and inverting with respect to $\tilde \mu$, we get
\begin{align} \label{Pnks_2}
{\cal P}_{N,\kappa}(s) = \frac{N!}{Z_N} \int_{\Gamma} \frac{d \tilde \mu}{2\pi i}  \int_< dx_1 \cdots dx_N \, e^{\tilde \mu \left(s - \frac{1}{N}\sum_{i=N(1-\kappa)+1}^N x_i \right)} e^{-\left[\frac{N^2}{2}\sum_{i=1}^N x_i^2 - \alpha \, N\,\sum_{i\neq i} |x_i-x_j| \right]} \;,
\end{align}
where $\Gamma$ is a Bromwich contour going along the imaginary axis in the complex $\tilde \mu$ plane. Note that this expression (\ref{Pnks_2}) can also be
obtained by replacing the delta-function in Eq. (\ref{Pnks}) by its integral representation $\delta(x) = \int_{-\infty}^\infty \frac{dq}{2\pi}\, e^{i q\,x}$ followed by the change of variable $q = i \tilde \mu$. In order that all the terms inside the exponential are of the same order, we rescale $\tilde \mu = \mu \,N^3$ to get
\begin{align} \label{Pnks_3}
{\cal P}_{N,\kappa}(s) = \frac{N^3\, N!}{Z_N} \int_{\Gamma} \frac{d \mu}{2\pi i} \,e^{\mu N^3 s}\,  \int_< dx_1 \cdots dx_N \, e^{-\beta E_\mu[\{x_i\}]} \;,
\end{align}
where
\bea \label{Emu}
\beta E_{\mu}[\{ x_i\}] = \frac{N^2}{2} \sum_{i=1}^N x_i^2 - \alpha \,N\, \sum_{i\neq j} |x_i - x_j| + \mu N^2 \sum_{i=N(1-\kappa)+1}^N x_i \;.
\eea
We denote the $N$-fold integral over the $x_i$'s in Eq. (\ref{Pnks_3}) as 
\bea \label{Zmu}
Z_N(\mu) = N!\,\int_< dx_1 \cdots dx_N \; e^{-\beta E_\mu[\{x_i\}]} \;.
\eea
Thus $Z_N(\mu)$ can be interpreted as the partition function of the original gas but in the presence of a chemical potential $\mu$ (analogue of the grand-canonical ensemble), that enforces the constraint on the TLS. In particular, note that the unconstrained partition function $Z_N = Z_N(\mu = 0)$. Hence for $\mu \neq 0$, we have two
species of particles: $N-N' = (1-\kappa)\,N$ ``left'' particles and $N' = \kappa\,N$ ``right'' particles. All of them interact with each other via the pairwise linear Coulomb repulsion. However the external potentials felt by the two species are different. The left ones feel only the harmonic potential $N^2\,x^2/2$. However, the right ones, in addition to this harmonic potential, also feel a linear potential $\mu N^2 x$, since the chemical potential acts only on the right particles.

\begin{figure}[t]
\centering
\includegraphics[width = 0.7\linewidth]{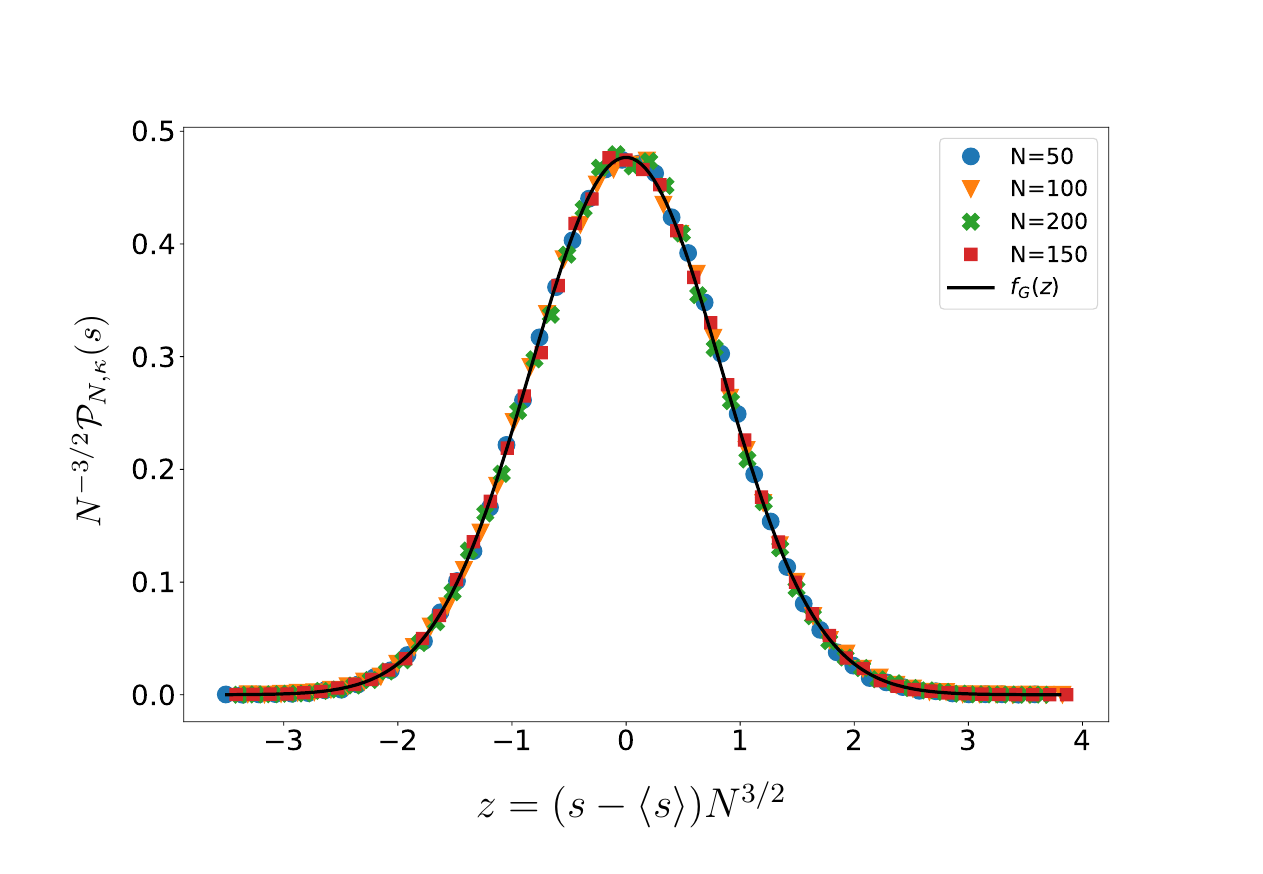}
\caption{Centered and scaled probability distribution $N^{-3/2}\,{\cal P}_{N,\kappa}(s)$ as a function of the scaling variable $z = N^{3/2}(s - \langle s \rangle)$ where $\langle s \rangle = 2 \alpha \kappa(1-\kappa)$, obtained by standard Monte-Carlo simulations with $\alpha = 1$ and $\kappa = 0.7$. The curves for different values of $N$ (symbols) collapse onto a single scaling function $f_G(z) = \e^{-z^2/(2 \kappa)}/\sqrt{2 \pi \kappa}$ (see Eqs.~(\ref{typ_Gaussian}) and (\ref{Gaussian_form})), indicated by the solid line. The agreement between the simulations and the theoretical prediction is excellent.}\label{Fig:typical}
\end{figure}

\vspace*{0.5cm}
\noindent{\it Typical fluctuations.} We have already seen in Eq. (\ref{s_av}) that the average value of the TLS $s$ is given by $\langle s \rangle = 2 \alpha \kappa(1-\kappa)$. 
The random variable $s$ fluctuates around this mean value.  To determine the scale of the typical fluctuations around $\langle s \rangle$ for large $N$, we set $s = \langle s \rangle + N^{-\varphi}\,z$ where $z$ is an $N$-independent random variable of order $O(1)$ and the exponent $\varphi$ is to be determined. Starting from the exact expression for ${\cal P}_{N,\kappa}(s)$ in Eq. (\ref{Pnks_3}), valid for arbitrary $N$, we take the large $N$ limit and show in \ref{sec:typical} that $\varphi = 3/2$ and moreover the distribution of $z$ is given by a pure Gaussian $f_G(z) = e^{-z^2/(2\kappa)}/\sqrt{2 \pi \kappa}$, as announced in Eq. (\ref{Gaussian_form}). In Fig. \ref{Fig:typical} we compare our analytical prediction for the scaling function $f_G(z)$ with Monte-Carlo simulations, finding an excellent agreement. This result also proves that the variance of $s$, for large $N$, is given by ${\rm Var}(s) \simeq \kappa/N^3$.

\vspace*{0.5cm}
\noindent{\it Atypically large fluctuations.} The fluctuations of size $|s - \langle s \rangle| \gg N^{-3/2}$ are not described by the Gaussian form discussed above. Instead, one needs to investigate the large deviation form of ${\cal P}_{N,\kappa}(s)$, which is achieved by a saddle-point method. Before proceeding to compute this large deviation form, it is useful to re-write the distribution 
${\cal P}_{N,\kappa}(s)$ in Eq. (\ref{Pnks_3}) as
\begin{eqnarray} \label{Pnks_4}
{\cal P}_{N,\kappa}(s) = \frac{N^3\, N!}{Z_N} \int_{\Gamma} \frac{d \mu}{2\pi i} \, \int_< dx_1 \cdots dx_N \, \e^{- S_{\mu}[\{x_i\}]} \;,
\end{eqnarray}
where 
\begin{align} 
S_\mu[\{x_i\}] =&  \frac{N^2}{2} \sum_{i=1}^N x_i^2 - \alpha \,N\, \sum_{i\neq j} |x_i - x_j| + \mu N^2 \left( \sum_{i=N(1-\kappa)+1}^N x_i -s \,N \right) \label{def_Smu} \\
=&  \; S_0[\{ x_i\}] + \mu N^2 \left( \sum_{i=N(1-\kappa)+1}^N f(x_i) -s \,N \right) \;, \label{def_Smu2}
\end{align}
where $f(x) = x$ and $S_0[\{ x_i\}] = \beta E[\{x_i\}]$ is the sum of the first two terms in Eq. (\ref{def_Smu}) denoting the ``bare'' scaled energy of the jellium model in Eq. (\ref{scaledE_2}). Although here $f(x) = x$ is very simple, the following analysis actually holds for arbitrary $f(x)$ (i.e., for general linear statistics).

We now perform the $(N+1)$-fold integral in Eq.~(\ref{Pnks_4}) by a saddle-point method for large $N$. Differentiating with respect to $x_i$'s and with respect to $\mu$ gives the saddle-point equations
\begin{align} 
\frac{\partial S_{\mu}}{\partial x_i} &= 0 \Longrightarrow \frac{\partial S_0}{\partial x_i} \Big \vert_{x_i^*}+ \mu N^2 f'(x_i^*) = 0  \label{sp_eq1} \\
\frac{\partial S_{\mu}}{\partial \mu} &= 0 \Longrightarrow \sum_{i=N(1-\kappa)+1}^N x_i^* = s \,N \;,\label{sp_eq2} 
\end{align}
where we denote the saddle-point configuration by $x_i^*$. The saddle-point equation (\ref{sp_eq2}), obtained by minimizing with respect to $\mu$, just gives the constraint that the TLS has a given value $s$. Ideally, we should denote the saddle-point value of $\mu$ by $\mu^*$ -- however, to keep the notation light, we will denote $\mu^*$ by $\mu$ and also 
suppress the explicit $\mu$-dependence in $x_i^*$. Substituting the values of $x_i^*$ in Eq. (\ref{sp_eq2}) determines $\mu$ as a function of $s$. Injecting the saddle-point solution $x_i^*$ in Eq. (\ref{def_Smu2}), one gets the saddle-point action $S_\mu[\{ x_i^*\}]$ as
\begin{align} \label{Smustar}
S_\mu[\{ x_i^*\}] = S_0[\{x_i^* \}] \simeq N^3 \psi_\kappa(s) \;.
\end{align}
The first equality follows upon using the second saddle-point equation (\ref{sp_eq2}) in (\ref{def_Smu2}). In addition, we will see that the saddle-point action scales as $N^3$ for large $N$. Substituting this saddle-point action (\ref{Smustar}) in Eq. (\ref{Pnks_4}) and using the large $N$ behavior of $Z_N$ in Eq. (\ref{ZNstar}) gives the large deviation form
\begin{align} \label{phi_psi}
{\cal P}_{N, \kappa}(s) \simeq \e^{-N^3 \phi_{\kappa}(s)} \quad, \quad {\rm where} \quad \phi_\kappa(s) = \psi_{\kappa}(s) - \frac{2 \alpha^2}{3} \;.
\end{align}

To evaluate the saddle-point action, and hence the rate function $\psi_\kappa(s)$ in Eq. (\ref{Smustar}), we will use the following nice short-cut method, valid
for generic Coulomb gases~\cite{Grabsch2015,Cunden2016,Grabsch2}. Taking a derivative of Eq. (\ref{Smustar}) with respect to $s$, and using chain rule, we get
\begin{align} 
N^3 \frac{\partial \psi_{\kappa}(s)}{\partial s}  &= \sum_{i=1}^N \frac{\partial S_0}{\partial x_i^*} \frac{\partial x_i^*}{\partial s} = - N^2 \mu(s) \sum_{i=1}^N f'(x_i^*) \frac{\partial x_i^*}{\partial s}  = - N^2 \mu(s)  \frac{\partial}{\partial s} \left[ \sum_{i=1}^N f(x_i^*)\right] \label{Smustar_deriv1}\\
&= - N^2 \mu(s) \frac{\partial}{\partial s}\left(s N\right) = - N^3 \mu(s) \;. \label{Smustar_deriv2}
\end{align}
In establishing the second equality in Eq. (\ref{Smustar_deriv1}) we used the saddle-point equation (\ref{sp_eq1}) and we have also exhibited the explicit $s$-dependence of $\mu(s)$. 
Integrating back the relation (\ref{Smustar_deriv2}), we get, up to an arbitrary constant
$\psi_{\kappa}(s)  = - \int^s \mu(s') \, ds'$. Therefore, from Eq. (\ref{phi_psi}), we get the rate function $\phi_\kappa(s)$ as
\begin{align} \label{phikappamu}
\phi_{\kappa}(s) = -  \int^s \mu(s') \, ds'  \;,
\end{align}
up to an arbitrary constant. To fix this constant, we proceed as follows. For the unconstrained case, we have seen before that the saddle-point solution is given by a flat configuration [see Eq. (\ref{E_GS})]. For this configuration, the associated value of the TLS in the large $N$ limit is $s = 2 \alpha \kappa(1-\kappa)$ [see Eq. (\ref{s_av})]. In other words, if we set 
$s = 2 \alpha \kappa(1-\kappa)$, the associated saddle-point configuration is the unconstrained flat configuration. For this configuration, ${\cal P}_{N,\kappa}(s)$ must be of order $O(1)$, which indicates that
\begin{align} \label{phi_sbar}
\phi_\kappa(s = 2 \alpha \kappa(1-\kappa)) = 0 \;.
\end{align} 
 This thus fixes the undetermined constant in (\ref{phikappamu}) and we get
 \begin{align} \label{phi_k_final}
 \phi_\kappa(s) = - \int_{2 \alpha \kappa(1-\kappa)}^s \mu(s') \, ds' \;.
 \end{align}
Hence, if we know $\mu(s)$, we obtain the exact rate function just by integrating $\mu(s)$. 

Our program for the rest of the paper is as follows. In the next section, we first determine the saddle-point configuration $x_i^*$ for a fixed $\mu$. We will see that the saddle-point solution will have different shapes depending on $(\kappa, \mu)$, leading to the phase diagram in Fig. \ref{fig:phase_space1}.
In the next section, using this saddle-point configuration, we evaluate $\mu(s)$ using Eq. (\ref{sp_eq2}). Eliminating $\mu$ in favour of $s$, this gives us the phase diagram in the $(\kappa, s)$ plane in Fig. \ref{fig:phase_space2}. We then use this expression for $\mu(s)$ to compute the rate function in Eq. (\ref{phi_k_final}). We will see that, as we cross the boundaries in the $(\kappa,s)$ plane (e.g., by varying $s$ for fixed $\kappa$), the rate function $\phi_\kappa(s)$ becomes non-analytic as we cross the phase boundaries. In particular, the third derivative of the rate function exhibits a discontinuity, leading to a third-order phase transition.

\section{Minimum energy configuration in the grand-canonical ensemble}\label{sec:deriv}

As mentioned above, the constrained partition function in Eq. \eqref{Zmu}, for large $N$, is dominated by the configurations $\{x_i^*\}$
that minimize the energy $E_\mu[\{x_i\}]$ in Eq.~\eqref{Emu} for fixed $\kappa$ and $\mu$. To facilitate this minimization, it is convenient 
to re-write the energy using the ordering $x_1<x_2< \cdots< x_N$ as 
\bea \label{Emu_ord1}
\hspace*{-1cm}\beta E_{\mu}[\{x_i\}] &=& \frac{N^2}{2}\sum_{i=1}^N x_i^2 - 2 \alpha N \, \sum_{i>j}(x_i-x_j) + \mu N^2 \sum_{i=N(1-\kappa)+1}^N x_i \\
\hspace*{-1cm}&=&  \frac{N^2}{2}\sum_{i=1}^N x_i^2 - 2 \alpha N \sum_{i=1}^N(2i-N-1)\,x_i + \mu N^2 \sum_{i=N(1-\kappa)+1}^N x_i \;,
\eea
where we used the identity $\sum_{i>j}(x_i-x_j) = \sum_{i=1}^N(2i-N-1)x_i$, which is proved in~\ref{sec:appsum}. We note that, for future use, the energy can also
be expressed for arbitrary $\kappa$ and $\mu$ as
\bea \label{eq:mu_negative}
\beta E_{\mu}[\{x_i\}] = {\cal F}_{\mu}[\{x_i\}] + C_1 \;,
\eea
with
\bea \label{calF_mu}
{\cal F}_{\mu}[\{x_i\}] &=& \frac{N^2}{2}\Bigg [  \sum_{i=1}^{N-N'}\left(x_i-\left[\frac{2\alpha}{N}(2i-N-1)\right]\right)^2 \nonumber \\ 
&+&  \sum_{i=N-N'+1}^{N}\left(x_i-\left[\frac{2\alpha}{N}(2i-N-1)-\mu \right]\right)^2 \Bigg]  \;,
\eea
where $N' = \kappa\,N$ and $C_1$ is just a constant given by
\begin{align} 
&C_1=-2\alpha^2\sum_{i=1}^{N}(2i-N-1)^2 - \frac{N^2}{2}\sum_{i=N(1-\kappa)+1}^{N}\big[ \mu^2 - \frac{4\alpha}{N}(2i-N-1)\mu\big]  \nonumber \\
 & \quad= N^3 \left[ -\frac{2\alpha^2}{3}- \frac{\mu^2 \kappa}{2} + 2 \alpha \kappa(1-\kappa)\mu\right] + \frac{2\alpha^2}{3}\,N  \;.
\label{eq:const}
\end{align}
We will now fix $\kappa$ and vary $\mu$ in the $(\kappa, \mu)$ plane. We will see that the minimum energy configurations are
principally of two types depending on whether $\mu \leq 0$ or $\mu >0$. For $\mu \leq 0$, we will see that the density profile in
the minimum energy configuration has two disjoint supports with a gap between them -- we will call this phase $I$. In contrast, for $\mu >0$, the gap disappears, there is only a single support but in addition, there is a delta peak inside the support coexisting with a uniform background density. In this second case $\mu>0$, the location
and weight of the delta peak varies with $\mu$ and $\kappa$, leading to four different sub-phases $II, IIIa, IIIb$ and $IV$ (see the phase diagram and
the associated density profiles in Fig. \ref{fig:phase_space1}). Below, we will discuss the two cases $\mu \leq 0$ and $\mu >0$ separately.

\subsection{The case $\mu \leq 0$: Phase $I$}

\begin{figure}[t]
\centering
\includegraphics[width=0.8\linewidth]{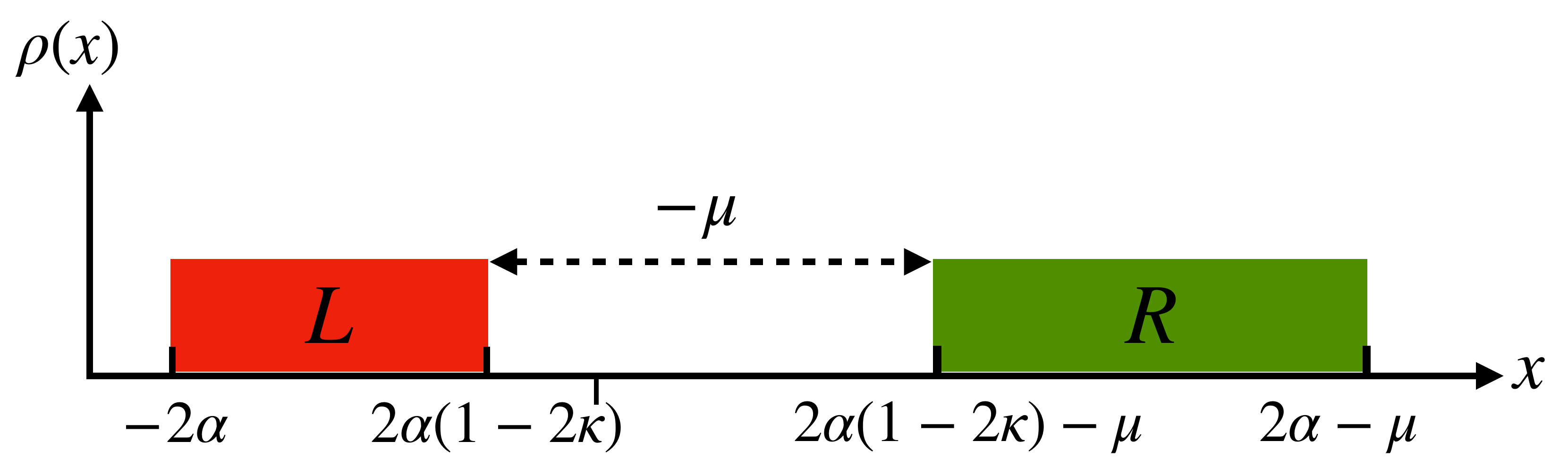}
\caption{The density profile associated to the minimum energy configuration for $\mu<0$. The density is flat over two disjoint intervals, noted
by left ($L$) and right ($R$) which are separated by a gap $-\mu>0$ [see Eq. \eqref{rho_mu_neg}].}\label{Fig_gap}
\end{figure}

\begin{figure}[t]
	\centering
	\includegraphics[width=0.8\textwidth]{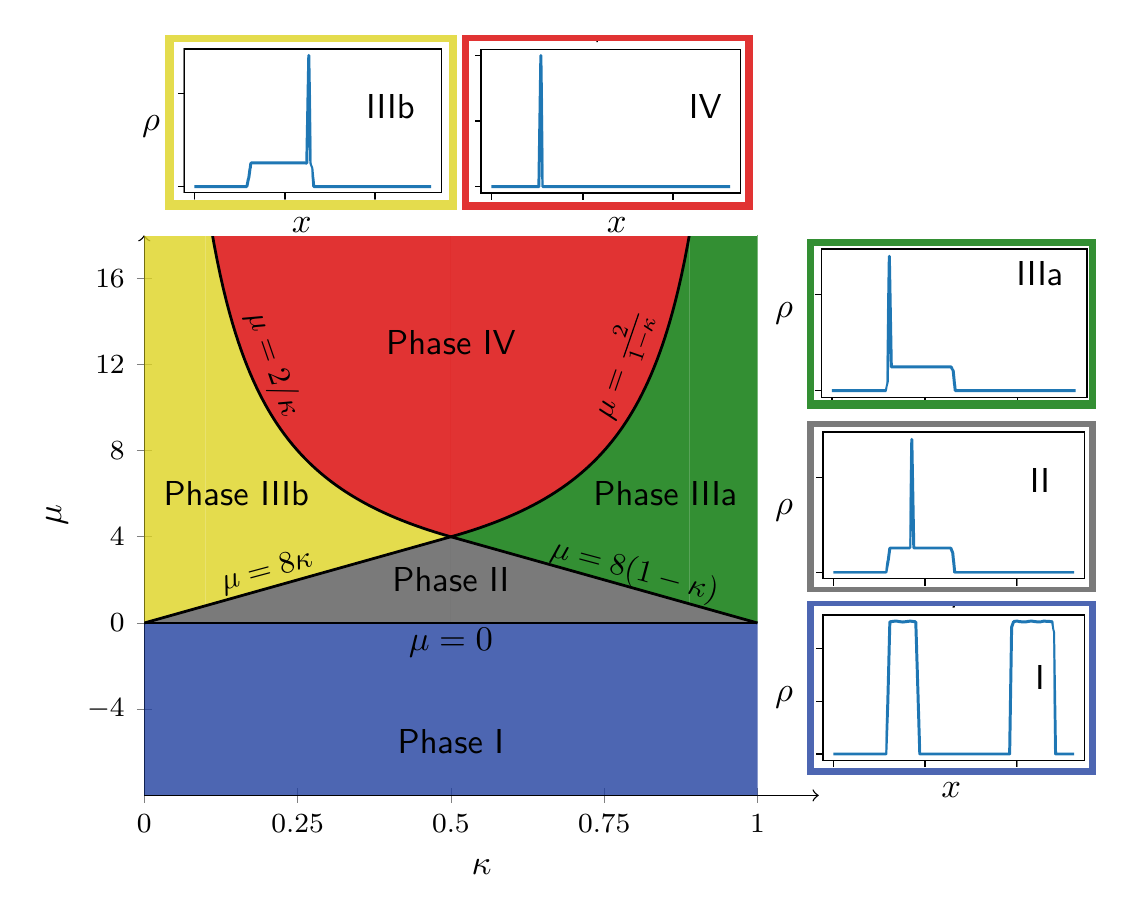}
	\caption{Phase diagram in the $(\kappa$-$\mu)$-plane for $\alpha =1$. We found five different phases for the jellium model: Phase $I$ (blue), phase $II$ (grey), phase $IIIa$ (green) phase $IIIb$ (yellow) and phase $IV$ (red) with different minimum energy configuration for a fixed $\mu$ (grand-canonical ensemble). For $0 < \kappa \leq 1/2$, the phase boundaries are given respectively $\mu=0$, $\mu = 8 \alpha \kappa$ and $\mu = 2\alpha/\kappa$ (from bottom to top). In contrast, for $1/2 \leq \kappa \leq 1$, the corresponding phase boundaries are given by $\mu = 0$, $\mu = 8 \alpha(1-\kappa)$ and $\mu = 2\alpha/(1-\kappa)$ (from bottom to top). On the upper and right edge of the figure we see the associated density profiles where the colour of the borders of the windows correspond to the colours of the phases.}
	\label{fig:phase_space1}
	\end{figure}

When $\mu \leq 0$, for arbitrary $0<\kappa\leq 1$, it is convenient to use the representation of the energy in Eq. \eqref{eq:mu_negative}. Note that since the constant $C_1$ is independent of $x_i$, it plays no role in minimizing the energy with respect to $x_i$'s. Hence, in Eq. \eqref{eq:mu_negative}, we just need to minimize ${\cal F}_\mu[\{x_i\}]$ given in Eq. (\ref{calF_mu}), which has two sums. In this case,
it turns out that each term of the two sums in Eq. \eqref{calF_mu} can be set to $0$ and the resulting configuration $x_i^*$ satisfies the ordering condition since $\mu \leq 0$. It is easy
to see that, since the energy is the sum of squares, this configuration satisfying the ordering is indeed the minimum energy configuration. This gives two blocks of solutions: 
the left species ($L$) corresponding to the first sum in Eq. \eqref{calF_mu} ranging from $i=1$ to $i=N-N'$ and the right species ($R$) associated to the second sum 
in Eq. \eqref{calF_mu} ranging from $i=N-N'+1$ to $N$. The solution then reads
\be
x_i^* \Big \vert_L =\frac{2\alpha}{N}(2i-N-1), \quad i = 1, 2, \cdots, (1-\kappa)\,N
\label{eq:eq_phaseIb}
\ee
and 
\be
	x_i^*\Big\vert_R =\frac{2\alpha}{N}(2i-N-1)-\mu, \quad i = (1-\kappa)N+1, \cdots, N \;.
	\label{eq:eq_phaseIa}
\ee
Taking the large $N$ limit, it then follows that the density of the particles in this minimum energy configuration
is again flat, but is supported over two disjoint intervals separated by a gap of length $-\mu >0$ (which thus satisfies
the ordering condition for $\mu<0$):
\begin{align}\label{rho_mu_neg}
\rho(x)=\left\{
\begin{array}{ll}
	\frac{1}{4 \alpha} \;, \; \quad -2\alpha<x<2\alpha(1-2\kappa),\\
	\frac{1}{4 \alpha} \;,\; \quad 2\alpha(1-2\kappa)-\mu<x<2\alpha-\mu,\\
	0 \;,\; \quad \textnormal{elsewhere}.
\end{array} \right.
\end{align}
This density profile is shown in Fig. \ref{Fig_gap}. This solution is valid for all $\mu \leq 0$ and arbitrary $0< \kappa \leq 1$. This is denoted by phase $I$ in the 
phase diagram in the $(\kappa, \mu)$ plane in Fig.~\ref{fig:phase_space1}, with the associated density profile also shown in the inset (with blue border). When $\mu$ exceeds $0$, this is no longer an acceptable solution, as it violates the
ordering property (the left $L$ and the right $R$ overlap) and one needs to find out the correct minimum energy configuration, as discussed below in details.

\subsection{The case $\mu >0$}

\begin{figure}[t]
	\centering
	\includegraphics[width=0.6\textwidth]{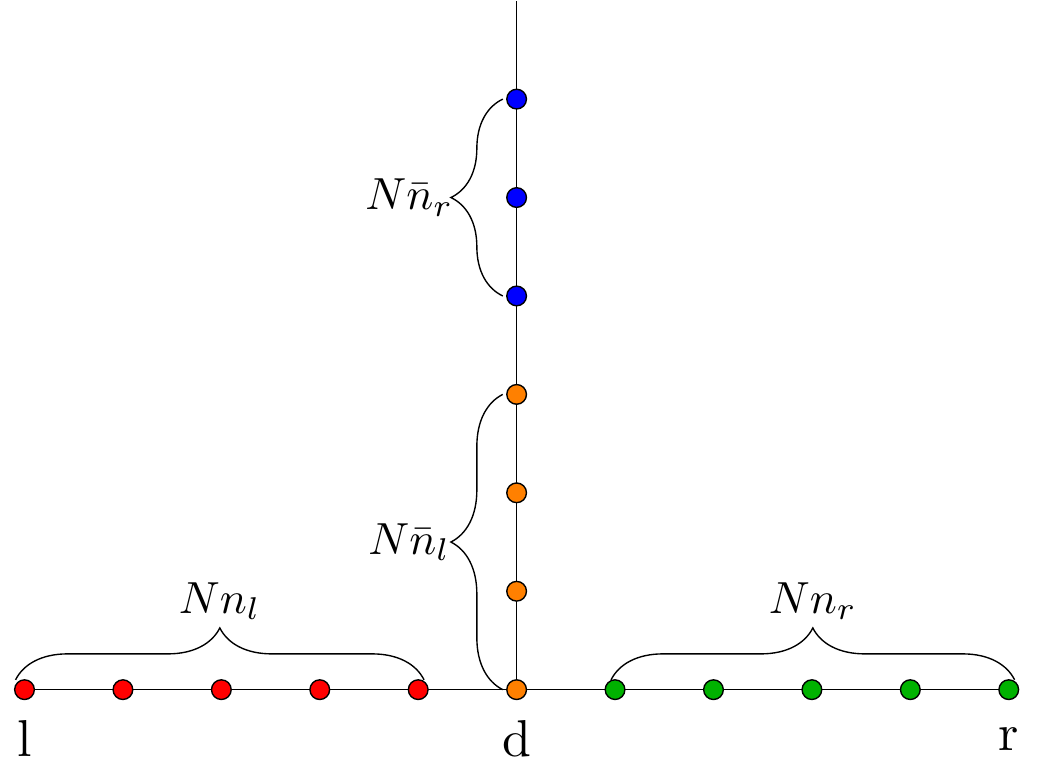}
	\caption{A schematic representation of the saddle configuration of the charge densities for $\mu > 0$. The ``tower'' at the location $d$ represents a delta-function which contains $N\,\bar{n}_l$ orange particles and $N \bar{n}_r$ blue particles. The number of red particles to the left of the delta-function is $N n_l$, while the number of green particles to the right of the delta function is denoted by $N n_r$. We denote the locations of the rightmost and the leftmost particles respectively by $r$ and $l$.}
	\label{fig:model}
\end{figure}

We have seen that when $\mu \to 0^-$ from phase $I$, the gap between the disjoint supports in Fig. \ref{Fig_gap} vanishes. This suggest 
that when $\mu$ exceeds $0$, the two species of charges may overlap and lead to a density profile shown schematically in Fig. \ref{fig:model} 
where a fraction of particles from the left and another fraction from the right pile up at a single point, leading to a delta-function in the density 
profile. In addition, there is a uniform background charge density both on the left and the right of the delta-peak, belonging respectively to
the two species $L$ and $R$. In fact, this background uniform density is expected to be the same as the unconstrained case, namely $\rho(x) = 1/(4 \alpha)$, and can be shown by minimizing the energy at any point in the bulk~\cite{Dhar2017,Dhar2018}. We can parametrize such a configuration by seven parameters (see Fig. \ref{Fig_gap}): 
\begin{itemize}
	\item the position of the leftmost particle: $l$,
	\item the position where the delta-function occurs, i.e., where fractions of charges from the left and right pile up: $d$,
	\item the position of the rightmost particle: $r$,
	\item the fraction of charges that are on the left of $d$: $n_l$,
	\item the fraction of charges that are on the right side of $d$: $n_r$,
	\item the fraction of particles at the point $d$ that came from the left: $\bar{n}_l$,
	\item the fraction of particles at the point $d$ that came from the right: $\bar{n}_r$.
\end{itemize}
Of course, these seven parameters are not independent, since there are relations between them. Since $\kappa = N'/N$ (the fraction of right species $R$) is fixed, we have the following relations
\begin{align} \label{cond1}
n_r + \bar{n}_r = \kappa \quad,\quad 
n_l + \bar{n}_l =1-\kappa \;,
\end{align}
which clearly implies $0 \leq \bar{n}_l \leq 1 - \kappa$ and $0 \leq \bar{n}_r \leq \kappa$. Another pair of relations may be obtained as follows
\begin{align}
\frac{1}{4\alpha}(d-l)=n_l \;,
\quad \frac{1}{4 \alpha}(r-d)=n_r \;.
\label{cond2} 
\end{align}
The first one follows from the fact that the fraction of charges $n_l$ in the bulk to the left of the delta-peak is just the distance $d-l$ times the uniform bulk density $1/(4 \alpha)$. Similarly, the second relation follows from the same argument applied to the right of the delta-peak. Since we have seven parameters and four relations between them, we have only three independent parameters left, which we choose to be $l, \bar{n}_l, \bar{n}_r$. Substituting this ansatz (see Fig. \ref{fig:model}) for the minimum energy configuration in the expression for the energy in Eq. (\ref{Emu_ord1}) and simplifying, one finds to leading order for large $N$
 
\begin{align} \label{E3param}
\frac{\beta E_{\mu}}{N^3}\simeq & 
-2 {\bar{n}_l}^2 (-4 \kappa +l+2)+\frac{8 {\bar{n}_l}^3}{3}-4 {\bar{n}_l} \left(2
   {\bar{n}_r}^2-4 {\bar{n}_r} \kappa +\kappa  (l+\mu +2)\right) +\frac{1}{2} l (l+4)+\frac{4}{3} \nonumber \\
&  \; +2 {\bar{n}_r}^2 (4 \kappa
   +l+\mu +2)-\frac{16 {\bar{n}_r}^3}{3}-4 {\bar{n}_r} \kappa  (l+\mu +2)+4 \kappa  \mu
   +\kappa  \mu  (l-2 \kappa )  \nonumber \\
 = &\;F(l, \bar{n}_l, \bar{n}_r)    
   \;.
\end{align}
The next step is to minimize this energy in Eq. \eqref{E3param} with respect to the three parameters $l, \bar{n}_l$ and $\bar{n}_r$. Taking derivatives of $F(l, \bar{n}_l, \bar{n}_r)$ with 
respect to these three variables, and setting them to zero, gives three equations
\begin{align} 
\frac{\partial F}{\partial l} &= 0 \Longrightarrow \quad l - 2 \alpha \left(-1+\bar{n}_l^2 - \bar{n}_r^2 + 2 \kappa(\bar{n}_l + \bar{n}_r) \right) +  \kappa \mu = 0  \label{mini1} \\
\frac{\partial F}{\partial \bar{n}_l} &=0  \Longrightarrow \quad l (\kappa+\bar{n}_l) + 2 \alpha\left( (1-\bar{n}_l)\bar{n_l} + \bar{n}_r^2 + \kappa - 2 \kappa(\bar{n}_l + \bar{n}_r)  \right) + \kappa \mu = 0  \label{mini2}\\
\frac{\partial F}{\partial \bar{n}_r} &=0  \Longrightarrow \quad (\bar{n}_r- \kappa)\left(l + \alpha(2-4 \bar{n}_l - 4 \bar{n}_r)+\mu\right) = 0  \label{mini3} \;.
\end{align}
The solutions to these three equations can be obtained using Mathematica. It turns out that there are four distinct physically acceptable solutions, once we use the two constraints 
$0 \leq \bar{n}_l \leq 1 - \kappa$ and $0 \leq \bar{n}_r \leq \kappa$. 

\subsection{Phase $II$}

In this case the solution is given, for all $0 < \kappa \leq 1$, by
\begin{align} \label{sol_II}
l = - 2 \alpha \quad, \quad \bar{n}_l = \frac{\mu}{8 \alpha} \quad, \quad \bar{n}_r = \frac{\mu}{8 \alpha} \;.
\end{align}
These two constraints $0 \leq \bar{n}_l \leq 1 - \kappa$ and $0 \leq \bar{n}_r \leq \kappa$ indicate that this solution (\ref{sol_II}) is valid for
\begin{align}\label{bound_II}
0< \mu \leq 8\alpha \min\left( \kappa, 1 - \kappa\right) \quad {\rm with} \quad 0 < \kappa \leq 1  \;.
\end{align}
These two bounds define the phase boundary of phase $II$ (grey triangle) in Fig. \ref{fig:phase_space1} (where we set $\alpha=1$). The upper bound is clearly symmetric 
around $\kappa = 1/2$, defining the two top arms of the grey triangle in Fig. \ref{fig:phase_space1}: $\mu = 8 \alpha \kappa$ (for $0 \leq \kappa \leq 1/2$) and $\mu = 8 \alpha(1-\kappa)$ (for $1/2\leq \kappa \leq 1$). The density profile corresponding to the parameters in Eq. \eqref{sol_II} is shown in the inset (with grey border) of Fig. \ref{fig:phase_space1}. 

\subsection{Phase IIIa}

Here, we set $1/2 \leq \kappa \leq 1$ and the solution is given by
\begin{align} \label{sol_IIIa}
l = - 2 \alpha - \mu + 2 \sqrt{2\alpha\,\mu(1-\kappa)} \quad, \quad \bar{n}_l = 1-\kappa \quad, \quad \bar{n}_r = \frac{-2 \alpha(1-\kappa) + \sqrt{2\alpha\mu(1-\kappa)}}{2 \alpha}\;,
\end{align}
with the condition that $2 \alpha(1-\kappa)< \mu < {2\alpha}/{(1-\kappa)}$, in order that $\bar{n}_l$ and $\bar{n}_r$ satisfy the inequalities $0 \leq \bar{n}_l \leq 1 - \kappa$ and $0 \leq \bar{n}_r \leq \kappa$. This solution is valid in the region
\begin{align}\label{boundIIIa}
8 \alpha(1-\kappa) < \mu < \frac{2\alpha}{1-\kappa} {\quad} {\rm with} \quad \frac{1}{2} \leq \kappa \leq 1 \;.
\end{align}
Note that the lower bound does satisfy the inequality $2 \alpha(1-\kappa)< \mu$ mentioned above. These two bounds in Eq. \eqref{boundIIIa} provide the loci of the phase boundaries of phase $IIIa$ shown by the green region of Fig. \ref{fig:phase_space1}, with $1/2 \leq \kappa \leq 1$ (and $\alpha$ set to unity). The density profile corresponding to the parameters in Eq. \eqref{sol_IIIa} is shown in the inset (with green border) of Fig. \ref{fig:phase_space1}.

\subsection{Phase $IIIb$}

Here, we set $0 \leq \kappa \leq 1/2$ and the solution is given by
\begin{align}\label{solIIb}
l = -2 \alpha \quad, \quad \bar{n}_l = \frac{-2 \sqrt{\alpha}\, \kappa + \sqrt{2 \mu \kappa}}{2 {\sqrt{\alpha}}} \quad, \quad \bar{n}_r = \kappa \;,
\end{align}
with the condition that $2 \alpha \kappa < \mu < {2\alpha}/{\kappa}$, in order that $\bar{n}_l$ and $\bar{n}_r$ satisfy the inequalities $0 \leq \bar{n}_l \leq 1 - \kappa$ and $0 \leq \bar{n}_r \leq \kappa$. This solution is valid in the region
\begin{align}\label{boundIIIb}
8 \alpha\,\kappa < \mu < \frac{2\alpha}{\kappa} {\quad} {\rm with} \quad 0 \leq \kappa \leq \frac{1}{2} \;.
\end{align}
These two bounds in Eq. \eqref{boundIIIb} provide the loci of the phase boundaries of phase $IIIb$ shown by the yellow region of Fig. \ref{fig:phase_space1}, with $0\leq \kappa \leq 1/2$ (and $\alpha$ set to unity).
In fact the region $IIIb$ is just a mirror image of the region $IIIa$ around $\kappa = 1/2$ (see Fig. \ref{fig:phase_space1}). The density profile corresponding to the parameters in Eq. \eqref{solIIb} is shown in the inset (with yellow border) of Fig. \ref{fig:phase_space1}.

\subsection{Phase $IV$}

Here the solution is given by
\begin{align}\label{solIV}
l = - \kappa\, \mu \quad, \quad \bar{n}_l = 1 - \kappa \quad, \quad \bar{n}_r = \kappa \;.
\end{align}
This solution corresponds to the case all the particles are in the delta-peak, as indicated in the inset (with red border) of Fig. \ref{fig:phase_space1}. This solution 
is valid in the region 
\begin{align} \label{boundIV}
\mu \geq \frac{2 \alpha}{\min(\kappa, 1 - \kappa)} \;.
\end{align}
This bound provides the phase boundary of phase $IV$ in Fig. \ref{fig:phase_space1} shown by the red colour, where again $\alpha = 1$ for convenience.  

\section{Phase diagram in the $(\kappa,s)$ plane}\label{sec:phdiag}

In this section, we start with the saddle-point configuration $\{x_i^*\}$ determined in the previous section for fixed $\kappa$ and $\mu$. We evaluate $s = \frac{1}{N}\sum_{i=N(1-\kappa)+1}^N x_i^*$ [see Eq. (\ref{sp_eq2})] for a given $\mu$. Inverting this relation gives us the desired $\mu(s)$ as a function of $s$ and also translates the phase boundaries from the $(\kappa, \mu)$ plane to the $(\kappa, s)$ plane, shown in Fig. \ref{fig:phase_space2}. We consider below the cases $\mu \leq 0$ and $\mu> 0$ separately.

\subsection{The case $\mu \leq 0$: phase $I$}

In this case, the saddle-point configuration has two disjoint supports with flat densities as in Eq. (\ref{rho_mu_neg}) and shown in Fig. \ref{Fig_gap}. 
Using Eq. (\ref{eq:eq_phaseIa}) for $x_i^*$, we get
\begin{align} \label{sphaseI}
s = \frac{1}{N}\sum_{i=N(1-\kappa)+1}^N x_i^* = \frac{2\alpha}{N^2}  \sum_{i=N(1-\kappa)+1}^N\left(2i-1 - \left(1+\frac{\mu}{2 \alpha} \right)N\, \right) \;.
\end{align}
Evaluating the sum explicitly and taking the large $N$ limit, we get
\begin{align} \label{sphase1large}
s \simeq {2 \alpha}\left( \left(1 - \frac{\mu}{2\alpha} \right)\kappa - \kappa^2 \right) \;.
\end{align}
Inverting this relation, we find
\begin{align} \label{musI}
\mu(s) = 2 \alpha(1-\kappa) - \frac{s}{\kappa} \;.
\end{align}
Since in this phase $\mu \leq 0$, this translates to the region
\begin{align}
\label{boundaryI}
s \geq 2\alpha\kappa(1- \kappa)
\end{align}
shown by the blue colour in the $(\kappa, s)$ plane in Fig. \ref{fig:phase_space2}. The phase boundary $\mu = 0$ in the $(\kappa, \mu)$ plane in Fig. \ref{fig:phase_space2}
then translates to the phase boundary 
\begin{align} \label{skappaI}
s(\kappa) = 2 \alpha \kappa(1-\kappa) \;. 
\end{align}

\subsection{The case $\mu > 0$}  

In this case, the saddle-point configuration $\{x_i^*\}$ is shown schematically in Fig. \ref{fig:model} and is characterised by seven parameters, of which only three are independent and we choose them to be $\{l, \bar{n}_l, \bar{n}_r\}$ as before. The TLS for such a configuration can be expressed as
\beal \label{TLS_mupos}
s = \frac{1}{N}\sum_{i=N(1-\kappa)+1}^N x_i^*  \;,
\end{align}
where the sum runs over the 'right' species of particles. These right particles are of two types: $N \bar{n}_r$ of them are in the delta-peak, i.e. they all stay at the same position $d$ and $N n_r$ of them that are located at equidistant points (with separation $4 \alpha/N$) to the right of the delta-peak. Hence the sum in Eq. (\ref{TLS_mupos}) can be split into two terms
\beal \label{TLS_mupos2}
s =  \frac{1}{N} \sum_{i=1}^{N\bar{n}_r} d + \frac{1}{N} \sum_{j=1}^{N\,n_r}\left(d+\frac{4\alpha}{N}j \right) = d(n_r + \bar{n}_r) + {2\alpha} n_r\left(n_r + \frac{1}{N} \right) \;.	
\end{align}	
For large $N$, dropping the $1/N$ term in (\ref{TLS_mupos2}) and using the relation $n_r + \bar{n}_r = \kappa$ we get
\beal 	\label{TLS_mupos3}
s \simeq d \,\kappa + 2 \alpha \left( \kappa - \bar{n}_r\right)^2 \;.
\end{align}
Furthermore, the location of the delta-peak $d$ can be expressed in terms of $l$ and $\bar{n}_l$ as follows
\beal \label{eq:d}
d = l + \frac{4\alpha}{N} \, N n_l = l + 4 \alpha(1- \kappa - \bar{n}_l) \;,
\end{align}
where we have used $n_l + \bar{n}_l = 1-\kappa$ [see Eq. (\ref{cond1})]. Hence, finally, in terms of the three independent parameters $\{l, \bar{n}_l, \bar{n}_r\}$, we have
\beal 	\label{TLS_mupos4}
s \simeq \kappa(l + 4 \alpha(1-\kappa-\bar{n}_l)) + 2 \alpha(\kappa-\bar{n}_r)^2  \;.
\end{align}

\subsection{Phase $II$}

In this case, the saddle-point solution for $\{l, \bar{n}_l, \bar{n}_r\}$ is given in Eq. (\ref{sol_II}). Substituting these values in Eq. (\ref{TLS_mupos4}) we find
\beal \label{s_phaseII}
s \simeq \kappa\left( 2 \alpha(1-2\kappa) - \frac{\mu}{2}\right) + 2 \alpha \left(\kappa - \frac{\mu}{8\alpha} \right)^2 \;.
\end{align}
To obtain $\mu$ as a function of $s$, we invert this relation (\ref{s_phaseII}), which amounts to solve a quadratic equation for $\mu$. This gives a priori two roots
\beal \label{two_roots}
\mu(s) = 4 \left( 4 \alpha \kappa \pm \sqrt{4 \alpha^2 \kappa^2 - 2 \alpha(2 \alpha\kappa(1-4 \kappa)-s)} \right) \;.
\end{align}
Of the two, only one (with the negative sign) satisfies the condition that $\mu(s) \to 0$ when $s$ approaches the phase boundary $s \to 2 \alpha \kappa(1-\kappa)$. Hence this gives
the unique function $\mu(s)$ in the phase $II$
\beal \label{mu_phaseII}
\mu(s) = 4 \left( 4 \alpha \kappa - \sqrt{4 \alpha^2 \kappa^2 - 2 \alpha(2 \alpha\kappa(1-4 \kappa)-s)} \right) \;.
\end{align}
To get the other phase boundaries of phase $II$, we note that in the $(\kappa, \mu)$ plane in Fig. \ref{fig:phase_space1}, the phase boundaries are given by $\mu = 8 \alpha \min(\kappa, 1-\kappa)$. For $0 \leq \kappa \leq 1/2$, using $\mu = 8 \alpha \kappa$ in Eq. (\ref{s_phaseII}), we get the phase boundary $s(\kappa) = 2 \alpha \kappa(1-4\kappa)$ between phase $II$ and phase $IIIb$ in Fig. \ref{fig:phase_space1}. For $1/2 \leq \kappa \leq 1$, using $\mu = 8\alpha(1-\kappa)$ in Eq. (\ref{s_phaseII}), we get $s(\kappa) = 2 \alpha(1-4\kappa)(1-\kappa)$, which describes the boundary between phase $II$ and phase $IIIa$ in Fig. \ref{fig:phase_space1}. Thus, summarising, the triangular grey region, corresponding to phase $II$ in Fig. \ref{fig:phase_space1} gets transformed into the grey region in Fig. \ref{fig:phase_space2}, whose upper boundary is $s(\kappa) = 2 \alpha \kappa(1-\kappa)$ for all $0 \leq \kappa \leq 1$, while the lower boundaries are described by the two curves
\begin{align} \label{s_phaseII_lower}
s(\kappa) = 
\begin{cases}
&2 \alpha \kappa(1-4 \kappa) \quad, \quad \quad \quad \;\; 0 \leq \kappa \leq 1/2 \\
&2 \alpha(1-\kappa)(1-4\kappa) \quad, \quad 1/2 \leq \kappa \leq 1 \;.
\end{cases}
\end{align}

\subsection{Phase $IIIa$}

Here we set $1/2 \leq \kappa \leq 1$ and the saddle-point solution for $\{l, \bar{n}_l, \bar{n}_r\}$ is given in Eq.~(\ref{sol_IIIa}).
Substituting these values in Eq. (\ref{TLS_mupos4}) we find
\begin{align} \label{s_phaseIIIa}
s = 2\alpha(1-\kappa) + (1-2\kappa)\mu - 2 \sqrt{2}(1-\kappa)\sqrt{\alpha \mu(1-\kappa)} \;.
\end{align}
We invert this relation and choose the root (with a negative sign again)	 such that $\mu(s) \to 8\alpha(1-\kappa)$ as  $s \to 2 \alpha(1-\kappa)(1-4 \kappa)$. This condition comes from the matching between phase $II$ and phase $IIIa$, along the curve $s(\kappa) = 2 \alpha(1-\kappa)(1-4\kappa)$ as in Eq.~(\ref{s_phaseII_lower}). This gives
\begin{align}\label{muIIIa}
\mu(s) = \frac{\alpha  (2-2 \kappa  (2 (\kappa -2) \kappa +3))-2 \sqrt{2} \sqrt{\alpha  (\kappa -1)^3 (2 \kappa  (\alpha  (\kappa -1)
   \kappa +s)-s)}-2 \kappa  s+s}{(1-2 \kappa )^2} \;,
\end{align}	
which is valid throughout phase $IIIa$ shown by the green colour in Fig. \ref{fig:phase_space2}. To obtain the phase boundaries of this region $IIIa$, we have already seen that the upper boundary is given by $s(\kappa) = 2 \alpha(1-\kappa)(1-4\kappa)$, that separates it from phase $II$. The lower boundary is obtained by setting $\mu = 2\alpha/(1-\kappa)$ [see Eq. (\ref{boundIIIa})] in Eq. (\ref{s_phaseIIIa}), which gives the boundary of the green region in Fig. \ref{fig:phase_space2} separating it from phase $IV$
\begin{align}\label{s_boundary_IIIa-IV}
s(\kappa) = - \frac{2\alpha \kappa^2}{1-\kappa} \quad, \quad {\rm for} \quad \frac{1}{2} \leq \kappa \leq 1 \;.
\end{align}
This curve diverges to $-\infty$ as $\kappa \to 1$, but in Fig. \ref{fig:phase_space2} we show this boundary only for $s \geq -3$.

\subsection{Phase $IIIb$}

Here we set $0 \leq \kappa \leq 1/2$, and the saddle-point solution for $\{l, \bar{n}_l, \bar{n}_r\}$ is given in Eq.~(\ref{solIIb}). Substituting these values in Eq. (\ref{TLS_mupos4}) we find
\begin{align} \label{s_phaseIIIb}
s = 2\kappa\left( \alpha - \sqrt{2 \alpha \mu \kappa }\right) \;.
\end{align}
Inverting this relation, we get
\begin{align} \label{muphaseIIIb}
\mu(s) = \frac{(2 \alpha \kappa - s)^2}{8 \alpha \kappa^3} \;.
\end{align}
The phase boundaries can be obtained as in the other cases. The boundary between phase $IIIb$ and phase $II$ is obtained by substituting $\mu = 8\alpha \kappa$ in Eq. (\ref{s_phaseIIIb}), giving $s(\kappa)= 2 \alpha \kappa(1-4\kappa)$. The boundary between phase $IIIb$ and phase $IV$ is obtained by setting $\mu = 2\alpha/\kappa$ [see Eq. (\ref{boundIIIb})] in Eq. (\ref{s_phaseIIIb}), which gives $s(\alpha) = -2 \alpha \kappa$. Thus summarising, the two boundaries of phase $IIIb$ (shown by the yellow region in Fig. \ref{fig:phase_space2}) are given by, for $0\leq \kappa \leq 1/2$
\begin{align} \label{bound_pahseIIIb}
s(\kappa) = 
\begin{cases}
&2 \alpha \kappa(1-4\kappa) \quad {\rm between} \quad IIIb \quad \& \quad II \\
& -2 \alpha \kappa \quad \quad  \quad \;\;\; {\rm between} \quad IIIb \quad \& \quad IV \;.
\end{cases}
\end{align}

\subsection{Phase $IV$} 

In this case,  the saddle-point solution for $\{l, \bar{n}_l, \bar{n}_r\}$ is given in Eq.~(\ref{solIV}). 
Substituting these values in Eq. (\ref{TLS_mupos4}) we find
\begin{align} \label{s_phaseIV}
s = - \mu \kappa^2 \;.
\end{align}
This gives
\begin{align} \label{mu_phaseIV}
\mu(s) = - \frac{s}{\kappa^2} \;.
\end{align}
The boundary between phase $IV$ and $IIIb$ is obtained by setting $\mu = 2 \alpha/\kappa$ with $0\leq \kappa \leq 1/2$ [see Eq. (\ref{boundIV})] in Eq. (\ref{s_phaseIV}). This gives $s(\kappa) = - 2 \alpha \kappa$ for $0\leq \kappa \leq 1/2$. Similarly, by setting $\mu = 2 \alpha/(1-\kappa)$ with $1/2 \leq \kappa \leq 1$ [see Eq. (\ref{boundIV})] in Eq. (\ref{s_phaseIV}), we recover the boundary between the phase $IV$ and $IIIa$, namely $s(\kappa) = -2 \alpha \kappa^2/(1-\kappa)$. Summarising, the two boundaries of phase $IV$ (shown as the red region in the $(\kappa, s)$ plane in Fig. \ref{fig:phase_space2}) are given by
\begin{align}\label{boundaryIV}
s(\kappa) = 
\begin{cases}
&- 2 \alpha \kappa \quad, \quad 0 < \kappa \leq 1/2  \\
&  -2 \alpha \kappa^2/(1-\kappa) \quad, \quad 1/2\leq \kappa \leq 1 \;.
\end{cases}
\end{align} 
This completes the description of the phase diagram in the $(\kappa, s)$ plane in Fig. \ref{fig:phase_space2}.

\section{Exact large deviation function $\phi_{\kappa}(s)$}\label{sec_large_dev}

In this section, we compute the exact large deviation function $\phi_{\kappa}(s)$ that describes the large $N$ behavior of ${\cal P}_{N, \kappa}(s) \simeq \e^{- N^3\phi_\kappa(s)}$. For this purpose,
we will use the relation in Eq. (\ref{phi_k_final}), which requires the knowledge of $\mu(s)$ in different phases of the phase diagram in Fig. \ref{fig:phase_space2}. In the previous section, we have computed $\mu(s)$ exactly in different parts of the $(\kappa,s)$ plane. In this section, we use these expressions of $\mu(s)$ to compute $\phi_\kappa(s)$, separately for $0 < \kappa \leq 1/2$ and $1/2 \leq \kappa \leq 1$. 

\subsection{The case $0 < \kappa \leq 1/2$}

We fix $0 <  \kappa \leq 1/2$ and we scan the phase diagram in Fig. \ref{fig:phase_space2} by decreasing $s$ continuously, starting from phase $I$. This way, we will encounter four different phases ($I$, $II$, $IIIb$ and $IV$), separated by three phase boundaries. These three phase boundaries for $0 \leq \kappa \leq 1/2$ were computed in the previous section and are summarised as follows
\begin{align} 
s_1(\kappa) &= 2 \alpha \kappa (1-\kappa) \quad, \quad \;\, {\rm between} \quad I \quad \& \quad II  \label{s1}\\
s_2(\kappa) & = 2 \alpha \kappa (1-4\kappa) \quad, \quad {\rm between} \quad II \quad \& \quad IIIb  \label{s2}\\
s_3(\kappa) & = - 2 \alpha \kappa   \quad, \quad \quad \quad \quad {\rm between} \quad IIIb \quad \& \quad IV  \label{s3}\;.
\end{align}
The expression for $\mu(s)$ in the four phases, computed in the previous section, are summarised below
\begin{align} \label{mu_fourphases1}
\mu(s) = 
\begin{cases}
&2 \alpha(1- \kappa) - \dfrac{s}{\kappa} \quad, \quad \hspace*{6.8cm} s_1(\kappa) \leq s \quad (I) \\
& \\
& 4 \left( 4 \alpha \kappa - \sqrt{4 \alpha^2 \kappa^2 - 2 \alpha(2 \alpha\kappa(1-4 \kappa)-s)} \right) \quad, \quad s_2(\kappa) \leq s \leq s_1(\kappa) \quad (II) \\
& \\
& \dfrac{(2 \alpha \kappa - s)^2}{8 \alpha \kappa^3}  \quad, \quad \quad  \hspace*{5.3cm} s_3(\kappa) \leq s \leq s_2(\kappa) \quad (IIIb) \\
& \\
& - \dfrac{s}{\kappa^2}  \quad, \quad  \quad  \hspace*{7.8cm} s \leq s_3(\kappa) \quad (IV)
\end{cases}
\end{align}
Identifying $s_1(\kappa) = 2 \alpha \kappa(1-\kappa)$, Eq. (\ref{phi_k_final}) reads 
\begin{align} \label{mu_s1}
\phi_\kappa(s) = - \int_{s_1(\kappa)}^s \mu(s')\, ds' \;.
\end{align}
We then carry out this integral using the different functional forms of $\mu(s)$ in Eq. (\ref{mu_fourphases1}) and obtain explicitly the rate function
\begin{align} \label{phi_k_lehalf}
\phi_{\kappa}(s) = 
\begin{cases}
&\dfrac{(s-2 \alpha  \kappa(1-\kappa ) )^2}{2 \kappa } \quad, \quad \hspace*{6.3cm} s_1(\kappa) \leq s \quad (I) \\
& \\
& 4 \left(-\frac{64}{3} \alpha ^2 \kappa ^3-8 \alpha ^2 (\kappa -1) \kappa ^2-4 \alpha \kappa \, s+\frac{2}{3} \sqrt{2 \alpha}  \; (2 \alpha  \kappa  (5 \kappa
   -1)+s)^{3/2}\right) \;, \\
&  \hspace*{9.2cm} s_2(\kappa) \leq s \leq s_1(\kappa) \quad (II) \\
& \\
& \dfrac{(2 \alpha  \kappa -s)^3}{24 \alpha  \kappa ^3} \quad, \quad \quad  \hspace*{5.3cm} s_3(\kappa) \leq s \leq s_2(\kappa) \quad (IIIb) \\
& \\
& \dfrac{2 \alpha ^2}{3}+\dfrac{s^2}{2 \kappa ^2} 
 \quad, \quad  \quad  \hspace*{7.cm} s \leq s_3(\kappa) \; \quad (IV)
\end{cases}
\end{align}
This function $\phi_\kappa(s)$ is plotted in the left panel of Fig. \ref{fig:LDF_integrated}. As $s$ decreases across the phase boundaries, the rate function $\phi_\kappa(s)$ and its first two derivatives with respect to $s$, namely $\phi_\kappa'(s)$ and $\phi_\kappa''(s)$, are continuous at all the three phase boundaries $s=s_1(\kappa)$, $s= s_2(\kappa)$ and $s=s_3(\kappa)$. However the third derivative is discontinuous at all the three boundaries and the jump discontinuities of the third derivatives $\phi_\kappa'''(s)$ at the three boundaries are given by
\begin{align} 
&\phi_\kappa'''(s \to s_1(\kappa)^+) - \phi_\kappa'''(s \to s_1(\kappa)^-) = \frac{1}{16 \alpha \kappa^3} \label{jump1} \\
&\phi_\kappa'''(s \to s_2(\kappa)^+) - \phi_\kappa'''(s \to s_2(\kappa)^-) = -\frac{1}{4 \alpha \kappa^3}  \label{jump2} \\
&\phi_\kappa'''(s \to s_3(\kappa)^+) - \phi_\kappa'''(s \to s_3(\kappa)^-) = -\frac{1}{4 \alpha \kappa^3} \label{jump3} \;.
\end{align} 
Thus the large deviation function exhibits third-order phase transitions with decreasing $s$ at each of the three phase boundaries $s=s_1(\kappa)$, $s=s_2(\kappa)$
and $s=s_3(\kappa)$ (see the left panel of Fig. \ref{fig:LDF_integrated} and also the upper left panel of Fig. \ref{fig:distr}). One can also verify that if the phase boundaries are traversed for a fixed $s$ by varying $\kappa$, one again finds a third-order phase transition at each phase boundary. Such third order phase transitions in the large deviation functions of full linear statistics of the form $(1/N)\sum_{i=1}^N f(x_i)$ in various long-range interacting systems, including log-gases \cite{Majumdar2014} as well as Coulomb gases in $d$-dimensions \cite{Dhar2017,Dhar2018,cunden2018universality}. It is interesting that we find third order phase transitions here, even for a truncated linear statistics.

\vspace*{0.5cm}
\noindent{\it Matching with typical fluctuations.} We first note that $s_1(\kappa) = 2 \alpha \kappa(1-\kappa)$ coincides with $\langle s \rangle$ in Eq. (\ref{s_av}). The large deviation $\phi_\kappa(s)$ describes the fluctuations of $s$, of order $O(1)$, around this mean value. If $|s - s_1(\kappa)| = |s - \langle s \rangle| \ll 1$, one expects that the large deviation form  
${\cal P}_{N, \kappa}(s) \simeq \e^{- N^3 \phi_\kappa(s)}$ must match with the form of the typical fluctuations described in Eqs. (\ref{typ_Gaussian}) and (\ref{Gaussian_form}). To verify that this is indeed the case, we analyse the rate function $\phi_\kappa(s)$ near its global minimum $s = s_1(\kappa) = \langle s \rangle$ [see the inset of Fig. \ref{fig:distr} and also the first two lines of Eq. (\ref{phi_k_lehalf})]. We find that $\phi_\kappa(s)$ has a quadratic form $\phi_\kappa(s) \simeq (s-\langle s \rangle)^2/(2 \kappa)$ as $s \to \langle s \rangle$. Consequently, the PDF of $s$ behaves, for large $N$ near $s = \langle s \rangle$ as 
\begin{align}\label{Gaussian_1}
{\cal P}_{N,\kappa}(s) \simeq \e^{-\frac{N^3}{2\kappa} (s-\langle s \rangle)^2} \quad,  \quad s \to \langle s \rangle \;.
\end{align}
This behavior thus matches perfectly with the form of the typical fluctuations given in Eqs. (\ref{typ_Gaussian}) and (\ref{Gaussian_form}).

\vspace*{0.5cm}
\noindent{\bf The limit $\kappa \to 0$ and the relation to the distribution of $x_{\max}$}. From the definition of $s = (1/N) \sum_{N-N'+1}^N x_i$ we see that if we set $N'=1$, i.e. $\kappa = 1/N$, then $s \to x_{\max}/N$. Then setting $\kappa = 1/N$ in Eqs. (\ref{s1})-(\ref{s3}), the phase boundaries become, to leading order for large $N$
\begin{align}
&s_1\left(\kappa \to \frac{1}{N}\right) = \frac{2 \alpha}{N} \label{s10} \\
&s_2\left(\kappa \to \frac{1}{N}\right) = \frac{2 \alpha}{N} = s_1\left(\kappa \to \frac{1}{N}\right) \label{s20} \\
&s_3\left(\kappa \to \frac{1}{N}\right) = - \frac{2 \alpha}{N} \label{s30}  \;.
\end{align}
Thus, to leading order, the two first boundaries $s_1(\kappa)$ and $s_2(\kappa)$ merge with each other. Hence in the $\kappa \to 0$ limit, we are left with only two boundaries in the $s_1(\kappa) \simeq 2\alpha/N$ and $s_3(\kappa) \simeq -2 \alpha/N$. Setting $\kappa = 1/N$ and $s = w/N$ (where $w$ denotes the value of $x_{\max}$), the rate function $\phi_\kappa(s)$ in Eq. (\ref{phi_k_lehalf}) then becomes  
\begin{align} \label{phi_k_lehalf0}
\phi_{\kappa = 1/N}(s=w/N) = 
\begin{cases}
&\dfrac{1}{2N}(w-2\alpha)^2 \quad, \quad \hspace*{3.7cm} 2 \alpha \leq w  \\
& \\
& \dfrac{(2 \alpha   - w)^3}{24 \alpha} \quad, \quad \quad  \hspace*{3.4cm} -2\alpha \leq w \leq 2\alpha  \\
& \\
& \dfrac{2 \alpha ^2}{3}+\dfrac{w^2}{2} 
 \quad, \quad  \quad  \hspace*{4.cm} w \leq -2\alpha  \;.
\end{cases}
\end{align}
If we now substitute this limiting expression of $\phi_{\kappa=1/N}(s=w/N)$ in the large deviation form ${\cal P}_{N,\kappa}(s) \simeq \e^{- N^3 \phi_\kappa(s)}$, we recover (i) for $w \leq 2 \alpha$, the right large deviation form of the PDF of $x_{\max}$ given in the third line of Eq. (\ref{Q_largeN}) with $\Phi_+(w) = (w-2\alpha)^2/2$ as in Eq.~(\ref{right_rf}) and (ii) for $w \leq 2 \alpha$, the left large deviation form given in the first line of Eq. (\ref{Q_largeN}) with $\Phi_-(w)$ as in Eq. (\ref{left_rf}). From our general discussion of the phase transition in Fig. \ref{fig:phase_space2}, we see that the two third-order phase transitions respectively at $s=s_1(\kappa=1/N) \simeq s_2(\kappa=1/N)$ and $s=s_3(\kappa=1/N)$ translate into two third-order phase transitions in the large deviation form of the PDF of $x_{\max}$ respectively at $w = 2 \alpha$  and $w = -2 \alpha$. In fact, these two third-order phase transitions in the large deviation of $x_{\max}$ were noticed and computed using a different method in Ref. \cite{Dhar2017, Dhar2018}. Here we nicely recover these results using the TLS formalism.

\subsection{The case $1/2\leq \kappa \leq 1$}

We now fix $1/2\leq \kappa \leq 1/2$ and we scan the phase diagram in Fig. \ref{fig:phase_space2} by decreasing $s$ continuously, starting from phase $I$. As in the previous case, we will encounter four different phases ($I$, $II$, $IIIa$ and $IV$), separated by three phase boundaries (only the third phase is different from the case $0 \leq \kappa \leq 1/2$). These three phase boundaries for $1/2 \leq \kappa \leq 1$ were computed in the previous section and are summarised as follows
\begin{align} 
\bar{s}_1(\kappa) &= 2 \alpha \kappa (1-\kappa) \quad, \quad \;\, \hspace*{1cm}{\rm between} \quad I \quad \& \quad II  \label{sb1}\\
\bar{s}_2(\kappa) & = 2 \alpha (1-\kappa) (1-4\kappa) \quad, \quad {\rm between} \quad II \quad \& \quad IIIa  \label{sb2}\\
\bar{s}_3(\kappa) & = - \frac{2 \alpha \kappa^2}{1-\kappa}   \quad, \quad \quad \quad \quad  \hspace*{0.7cm}{\rm between} \quad IIIa \quad \& \quad IV  \label{sb3}\;.
\end{align}
The expression for $\mu(s)$ in the four phases, computed in the previous section, are summarised below
\begin{align} \label{mu_fourphases1b}
\mu(s) = 
\begin{cases}
&2 \alpha(1- \kappa) - \dfrac{s}{\kappa} \quad, \quad \hspace*{6.8cm} \bar{s}_1(\kappa) \leq s \quad (I) \\
& \\
& 4 \left( 4 \alpha \kappa - \sqrt{4 \alpha^2 \kappa^2 - 2 \alpha(2 \alpha\kappa(1-4 \kappa)-s)} \right) \quad, \quad \bar{s}_2(\kappa) \leq s \leq \bar{s}_1(\kappa) \quad (II) \\
& \\
& \dfrac{2\alpha  (1-\kappa  (2 (\kappa -2) \kappa +3))+(1-2\kappa)s}{(1-2 \kappa )^2}  \\
&  -\dfrac{2 \sqrt{2} \sqrt{\alpha  (\kappa -1)^3 (2 \kappa  (\alpha  (\kappa -1)
   \kappa +s)-s)}}{(1-2 \kappa )^2} \;,\hspace*{0.8cm}\bar{s}_3(\kappa) \leq s \leq \bar{s}_2(\kappa) \quad (IIIa)\\
& \\
& - \dfrac{s}{\kappa^2}  \quad, \quad  \quad  \hspace*{7.8cm} s \leq \bar{s}_3(\kappa) \quad (IV) \;.
\end{cases}
\end{align}
Identifying, as before, $\bar{s}_1(\kappa) = 2 \alpha \kappa(1-\kappa)$, Eq. (\ref{phi_k_final}) reads 
\begin{align} \label{mu_s1b}
\phi_\kappa(s) = - \int_{\bar{s}_1(\kappa)}^s \mu(s')\, ds' \;.
\end{align}
We then carry out this integral using the different functional forms of $\mu(s)$ in Eq. (\ref{mu_fourphases1b}) and obtain explicitly the rate function
\begin{align} \label{phi_k_gehalf} 
\phi_{\kappa}(s) = 
\begin{cases}
&\dfrac{(s-2 \alpha  \kappa(1-\kappa ) )^2}{2 \kappa } \quad, \quad \hspace*{6.3cm} \bar{s}_1(\kappa) \leq s \quad (I) \\
& \\
& 4 \left(-\frac{64}{3} \alpha ^2 \kappa ^3-8 \alpha ^2 (\kappa -1) \kappa ^2-4 \alpha \kappa \, s+\frac{2}{3} \sqrt{2 \alpha}  \; (2 \alpha  \kappa  (5 \kappa
   -1)+s)^{3/2}\right) \;, \\
&  \hspace*{9.2cm} \bar{s}_2(\kappa) \leq s \leq \bar{s}_1(\kappa) \quad (II) \\
& \\
& \dfrac{4 \alpha ^2 (\kappa -1)^2 \left(8 \kappa ^4-8 \kappa ^3+4 \kappa -1\right)+12 \alpha  (\kappa -1) (2 \kappa -1) (2 (\kappa -1) \kappa +1)
   s}{6 (2 \kappa -1)^3}\\  
  & +\dfrac{s^2}{2 (2 \kappa -1)} +   \dfrac{4 \sqrt{2} \left(\alpha  (\kappa -1)^3 \left(2 \alpha  (\kappa -1) \kappa ^2+(2 \kappa -1) s\right)\right)^{3/2}}{3 \alpha  (\kappa
   -1)^3 (2 \kappa -1)^3} \;,\\
 & \hspace*{8.5cm} \bar{s}_3(\kappa) \leq s \leq \bar{s}_2(\kappa) \quad (IIIa) \\
& \\
& \dfrac{2 \alpha ^2}{3}+\dfrac{s^2}{2 \kappa ^2} 
 \quad, \quad  \quad  \hspace*{7.cm} s \leq \bar{s}_3(\kappa) \quad (IV) \;.
\end{cases}
\end{align}
This function $\phi_\kappa(s)$ is plotted in the right panel of Fig. \ref{fig:LDF_integrated}. As $s$ decreases across the phase boundaries, the rate function $\phi_\kappa(s)$ and its first two derivatives with respect to $s$, namely $\phi_\kappa'(s)$ and $\phi_\kappa''(s)$, are continuous at all the three phase boundaries $s=\bar{s}_1(\kappa)$, $s= \bar{s}_2(\kappa)$ and $s=\bar{s}_3(\kappa)$. However, as in the case of $0 \leq \kappa \leq 1/2$, the third derivative is discontinuous at all the three boundaries and the jump discontinuities of the third derivatives $\phi_\kappa'''(s)$ at the three boundaries are given by
\begin{align} 
&\phi_\kappa'''(s \to \bar{s}_1(\kappa)^+) - \phi_\kappa'''(s \to \bar{s}_1(\kappa)^-) = \frac{1}{16 \alpha \kappa^3} \label{jump11} \\
&\phi_\kappa'''(s \to \bar{s}_2(\kappa)^+) - \phi_\kappa'''(s \to \bar{s}_2(\kappa)^-) = -\frac{1}{4 \alpha (3\kappa-1)^3}  \label{jump21} \\
&\phi_\kappa'''(s \to \bar{s}_3(\kappa)^+) - \phi_\kappa'''(s \to \bar{s}_3(\kappa)^-) = -\frac{(1-\kappa)^3}{4 \alpha \kappa^6} \label{jump31} \;.
\end{align} 
Thus, as in the case of $0 \leq \kappa \leq 1/2$, here also  the large deviation function exhibits third-order phase transitions with decreasing $s$ at each of the three phase boundaries $s=\bar{s}_1(\kappa)$, $s=\bar{s}_2(\kappa)$ and $s=\bar{s}_3(\kappa)$, as shown in the right panel of Fig. \ref{fig:LDF_integrated}. Here also one can verify that if we traverse the phase boundaries by varying $\kappa$, while keeping $s$ fixed, that one encounters third-order transitions at the phase boundaries.  
As in the case $0 < \kappa \leq 1/2$, by investigating the quadratic behavior of $\phi_{\kappa}(s)$ near $s = \bar{s}_1(\kappa) = 2 \alpha \kappa(1-\kappa) = \langle s \rangle$, one finds
that for $1/2 \leq \kappa \leq 1$ also, the large deviation form matches smoothly with the typical form given in Eqs. (\ref{typ_Gaussian}) and (\ref{Gaussian_form}).

\vspace*{0.5cm}
\noindent{\bf The limit $\kappa \to 1$ and the relation to the distribution of the center of mass}. In the limit $\kappa \to 1$, the observable $s = (1/N) \sum_{i=1}^N x_i$ is
just the center of mass. In this case, setting $\kappa =1$ in Eqs. (\ref{sb1})-(\ref{sb3}), the phase boundaries become $\bar{s}_1(\kappa=1)= \bar{s}_2(\kappa=1)$ and $\bar{s}_3(\kappa = 1) \to - \infty$. Therefore, out of the four phases $I, II, IIIa$ and $IV$ in Fig.~\ref{fig:phase_space2}, only the two phases $I$ and $IIIa$ survive with $\bar{s_1}(\kappa=1) = 0$ denoting the boundary between them. Taking the limit $\kappa \to 1$ in phases $I$ and $IIIa$ in Eq. (\ref{phi_k_gehalf}), we get
\begin{align}\label{phik1}
\phi_{\kappa=1}(s) =
\begin{cases}
&\dfrac{s^2}{2} \quad, \quad s \leq 0 \\
&\\
&\dfrac{s^2}{2} \quad, \quad s \geq 0 \;. \\
\end{cases}
\end{align}
Indeed, there is no longer a phase transition in $\phi_{\kappa=1}(s)$ at $s=0$. Thus $\kappa=1$ is different from the $\kappa \to 1^-$ limit, where there are still three third-order phase transitions, with jump discontinuities in the third derivatives given in Eqs. (\ref{jump11})-(\ref{jump31}). Substituting $\phi_{\kappa=1}(s)$ from Eq. (\ref{phik1})
into the large deviation form ${\cal P}_{N, \kappa}(s) \simeq \e^{-N^3 \phi_\kappa(s)}$, we get a purely Gaussian distribution, for all $s$, which thus recovers the exact result for the distribution of the center of mass given in \ref{sec:appcofm}. 
	
\section{Monte-Carlo simulations} \label{sec:MC}

We would like to compare our analytical predictions for the distribution of the TLS with direct numerical simulations. One can start with the energy 
function given in Eq.~(\ref{scaledE_2}) and use a Metropolis dynamics to evolve the configurations of the charges. In the standard Metropolis dynamics,
from a given configuration $\{ x_i\}$, one proposes a small change $\{ \Delta x_i \}$ and accepts this change with a probability $\min\{1,\e^{-\beta \Delta E} \}$, where
$\Delta E$ is the change in energy as a result of the change in $\{x_i \}$ and $\beta$ is the inverse temperature. This dynamics satisfies detailed balance, 
which ensures that the system, at long times, reaches the equilibrium stationary state with the correct Boltzmann weight $\propto \e^{-\beta E}$. Once the system
has reached equilibrium, one can then compute the statistics of any observable, e.g., $s$ denoting the TLS. For the distribution of $s$, this method will only allow
to measure the typical fluctuations of $s$, of order $O(N^{-3/2})$ around its mean. However, we are interested here in the large deviations of $s$, i.e., the atypical large fluctuations of $s$,
of order $O(1)$. The probability of such large fluctuations is however extremely small, ${\cal P}_{N,\kappa}(s) \simeq \e^{-N^3 \phi_\kappa(s)}$. Configurations with such tiny probability
are very hard to sample via the standard Metropolis algorithm described above. Hence we need to adapt the algorithm to sample these rare atypical configurations, with a higher probability.
This is achieved by the so called importance sampling method \cite{Schawe2018,Nadal2010,Nadal2011,Hartmann2011,Hartmann2018,Banerjee2020,Mori2021}.

	We know that the average of our distribution is $\langle s \rangle =2\kappa\alpha(1-\kappa)$ and our goal is to explore regions to the right where $s>\langle s \rangle$ and regions to the left where $s<\langle s \rangle$. If we want to explore the region on the left from the average, we first pick an value $s^*<\langle s \rangle$  and only accept moves where $s_{\rm new}\leq s^{*}$. Similarly, for exploring the region to the right of $\langle s \rangle$ we can choose $s^*>\langle s \rangle$ and only accept moves with $s_{\rm new}\geq s^*$.\par 
	If we focus on the case $s<\langle s \rangle$, the main steps of the algorithm can be summarized as follows:
	\begin{itemize}
		\item Choose a initial configuration of $\{x_i\}$ that satisfies $s=\frac{1}{N}\sum_{i=(1-\kappa)N+1}^{N}x_i<s^*$.
		\item Propose a move of a particle and calculate $s_{\rm new}$ and $E_{\rm new}$. The new position of the particle is chosen as
		\be 
		\Delta x_i=d_{\max}(1-2n),
		\label{eq:delta}
		\ee
		where $n$ is drawn from a uniform distribution between $0$ and $1$. $d_{\max}$ is a real number and needs to be set in a way that the acceptance ratio is around $1/2$.
		\item If $s_{\rm new}>s^*$ we immediately reject the move.
		If $s_{\rm new}\leq s^*$ we accept the move with the probability
		\be 
		p=\min(\e^{-\beta(E_{\rm new}-E)}, 1)
		\label{eq:probability}
		\ee
		\item We repeat the previous step until we reach the equilibrium, this usually takes $10^7-10^8$ steps.
		\item We continue with the same process and sample $s$ every $100$ steps in order to construct $\mathcal{P}_{N,\kappa}(s)$.
	\end{itemize}
	
	\begin{figure}
		\centering
		\includegraphics[width=1\textwidth, height=7.5cm]{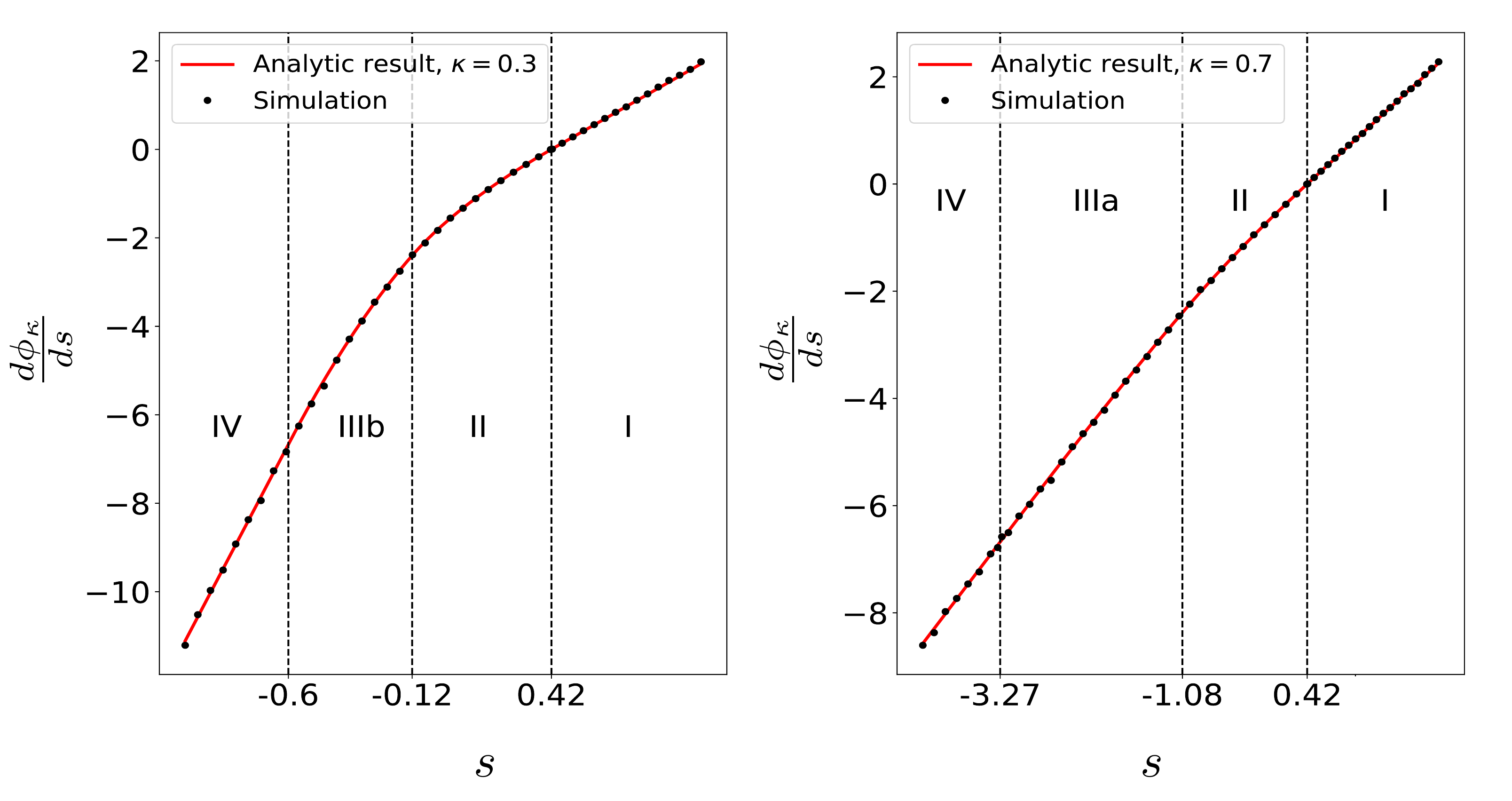}
		\caption{The derivative of the large deviation function $\phi_{\kappa}(s)$ obtained by simulation. The left panel shows the comparison between analytical solution and simulation for $\kappa=0.3$. The right panel depicts the analytical result and numerical data for $\kappa=0.7$. In both cases the black dotted vertical lines show the locations of the phase boundaries. For numerical simulations we used $N=50$ particles.}
		\label{fig:LDF_derivative}
	\end{figure}

	With this method we can explore small regions around $s^*$. In order cover a large interval of values $s$, we need to repeat the whole process for different $s^*$. \par 
	Another obstacle that arises is due to the additional restriction on the accepted moves. The result of above algorithm is not $\mathcal{P}_{N,\kappa}(s=S)$ but rather the conditional probability
	\be 
	\mathcal{P}_{N,\kappa}(s=S|s>s^*).
	\label{eq:conditional}
	\ee 
	The quantity we want to compute is therefore given by
	\be 
	\mathcal{P}_{N,\kappa}(s=S)=\mathcal{P}(s=S|s\geq s^*)P(s\geq s^*)
	\label{eq:consitional2}
	\ee
	Since $\mathcal{P}_{N,\kappa}(s)\simeq \e^{-N^3\phi_{\kappa}(s)}$, we need to take the logarithm and divide by $-N^3$ to obtain the large deviation function
	\be 
	\phi_{\kappa}(s) \simeq -\frac{1}{N^3}\ln\mathcal{P}_{N,\kappa}(s)=-\frac{1}{N^3}\left[\ln\mathcal{P}_{N,\kappa}\left(s=S|s\geq s^*\right)+\ln\mathcal{P}_{N,\kappa}\left(s\geq s^*\right)\right].
	\label{eq:conditional3}
	\ee
	We need to find a way to subtract the last constant term from the result in order to obtain the true large deviation function. Henceforth, what we have are many pieces of the result and they are centered around different values $s^*$. In addition each of them has an added unknown constant. To get rid off these constants we can look at the derivative of the large deviation function. Since different histograms centered around different $s^*$ do not overlap, we use a linear interpolation to compute ${d\phi_{\kappa}}/{d s}$.
	
	\begin{figure}
	\centering
	\includegraphics[width=\textwidth, height=7.5cm]{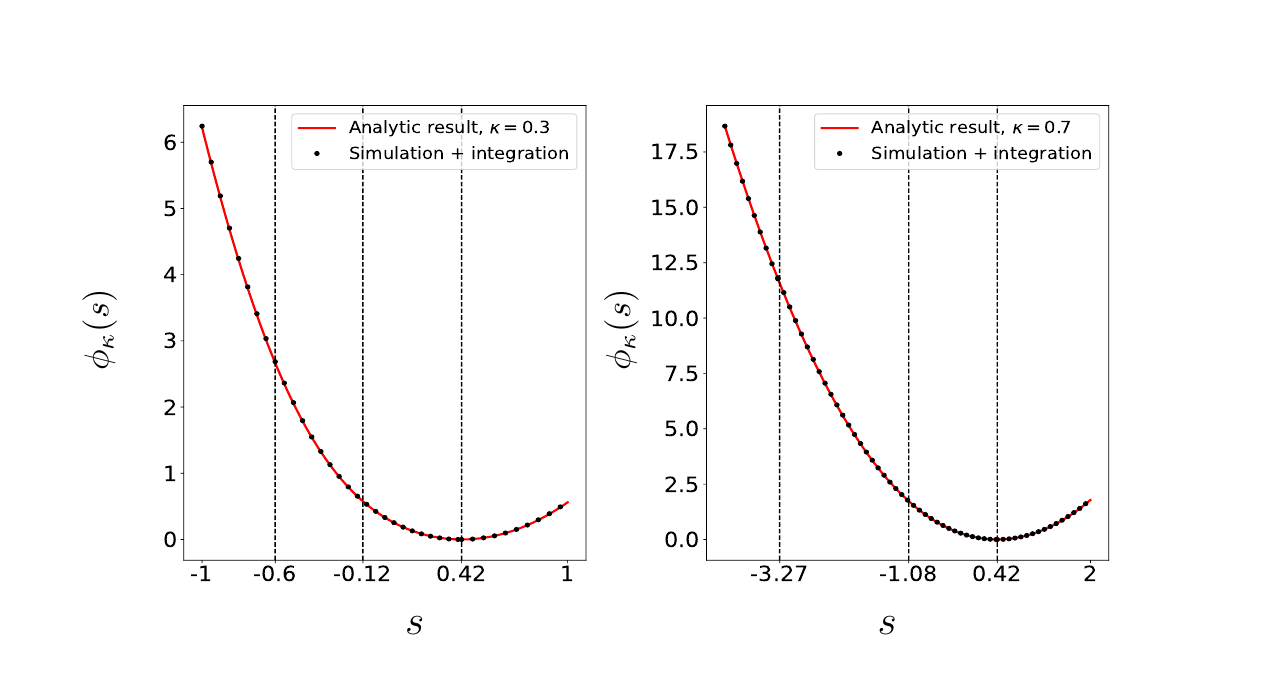}
	\caption{Comparison between the numerically computed large deviation function $\phi_\kappa(s)$ and its analytical predictions for $\kappa=0.3$ (left panel) and $\kappa=0.7$ (right panel).}
	\label{fig:LDF_integrated}
\end{figure}
	
	In Fig.~\ref{fig:LDF_derivative} we show the numerically obtained derivative $d\phi_{\kappa}(s)/ds$. In the left panel $\kappa$ is set to $\kappa = 0.3$ corresponding to the phase $IIIb$ in the phase diagram in Fig. \ref{fig:phase_space2}, while in the right panel $\kappa = 0.7$, corresponding to the phase $IIIa$. We used $50$ particles and we made approximately $10^8$ steps to reach the equilibrium. Then we sampled $s$ every $100$ moves.
	Finally we can numerically integrate the derivative to obtain the large deviation function and directly compare it with the analytical results. This can be seen in Fig.~\ref{fig:LDF_integrated}, showing excellent agreement between our analytical predictions and numerical simulations.

\section{Conclusion}\label{sec:conclu}

In this paper, we have studied the truncated linear statistics in the $1d$OCP with $N$ particles, where each particle is
subjected to a harmonic potential and they repel each other pairwise by the one-dimensional Coulomb interaction. 
Denoting by $x_i$'s the ordered positions of these particles $x_1 < x_2 < \cdots < x_N$, we focused 
on the $N'= \kappa N$ rightmost particles (with $0< \kappa \leq 1$) and studied the truncated 
linear statistics $s = \frac{1}{N}\sum_{i=(1-\kappa)N+1}^N x_i$. This observable $s$ is simply proportional to the position of the center of mass of the 
$\kappa N$ rightmost particles and interpolates in the two limits $\kappa \to 0$ and $\kappa \to 1$ respectively, between the position of the rightmost
particle $s=x_{\max}/N$ and the full center of mass.

We have computed analytically and numerically the distribution ${\cal P}_{N,\kappa}(s)$ 
in the large $N$ limit. The observable $s$ has its average value $\langle s \rangle = 2 \alpha \kappa(1-\kappa)$ and it fluctuates
around it. We have studied the probability of these fluctuations when they are typical, namely 
$s - \langle s \rangle$ of order $O(N^{-3/2})$, as well as when $s - \langle s \rangle$ is
of order $O(1)$ corresponding to atypically large fluctuations. Our results for ${\cal P}_{N,\kappa}(s)$ in the large $N$ limit 
can be summarised as follows
\begin{align} \label{summary}
{\cal P}_{N,\kappa}(s) \simeq
\begin{cases}
&N^{3/2} f_G\left( N^{3/2}(s-\langle s \rangle)\right) \quad, \quad |s-\langle s \rangle| = O(N^{-3/2}) \\
& \\
&\e^{-N^3 \phi_\kappa(s)} \quad, \quad \hspace*{2.7cm} |s-\langle s \rangle| = O(1) \;,
\end{cases}
\end{align}
where the scaling function $f_G(z) = \e^{-z^2}/(2 \kappa)/\sqrt{2 \pi \kappa}$ corresponding to typical fluctuations is a pure Gaussian.  
The rate function $\phi_\kappa(s)$ in the second line of Eq.~(\ref{summary}) describing the large deviation form is also computed exactly. 
Most interestingly, we have shown that there is a rich phase diagram in the $(\kappa, s)$ plane, with five
distinct phases where the rate function $\phi_\kappa(s)$ takes different functional forms (see Fig. \ref{fig:phase_space2}). In addition, we showed that, 
as one crosses the phase boundaries, while the rate function and its first two derivatives are continuous, the third derivative undergoes 
a jump, indicating a third order phase transition. We have also measured this rate function $\phi_\kappa(s)$ numerically, using a sophisticated importance sampling
method, adapted to compute the tails of a probability distribution very accurately. The numerical rate function matches very well with our analytical predictions. We have 
shown that in the limiting cases $\kappa \to 0$ and $\kappa \to 1$, our method, using this TLS formalism, perfectly recovers the large deviation form of the distributions of 
$x_{\max}$ and the center of mass, known previously by other methods. 

This paper focused on a specific TLS, namely the center of mass of the $\kappa N$ rightmost particles. It would be interesting to study the distribution of 
a more general TLS of the form $({1}/{N}) \sum_{i=(1-\kappa)N+1}^N f(x_i)$ where $f(x)$ is an arbitrary function. Furthermore, the repulsive potential between
a pair of particles in the $1d$OCP is of the one-dimensional Coulomb form, i.e. $\propto |x_i-x_j|$. A more general long-range model is the harmonically confined Riesz gas \cite{marcelriesz1938}
where the pairwise repulsion takes the form $\propto |x_i - x_j|^{-k}$ with $k > -2$ \cite{agarwal2019some,agarwal2019harmonically}. The $1d$OCP is a special case of the Riesz gas in the limit $k \to -1$. Similarly the Dyson's
log-gas corresponds to the $k \to 0$ limit and the classical Calogero-Moser model~\cite{Calogero75,Moser76} corresponds to $k=2$. Here, we 
we have studied the TLS only for $k=-1$. It would be interesting to extend these studies to the TLS of the general Riesz gas, with arbitrary interaction exponent $k > -2$.

\section*{Acknowledgments}
We thank F. Mori and C. Texier for useful discussions. This research was supported by ANR grant ANR-17-CE30-0027-01 RaMaTraF..

\appendix

\section{The distribution of the center of mass in $1d$OCP: the case~$\kappa = 1$}\label{sec:appcofm}

We start from the expression of ${\cal P}_{N, \kappa}(s)$ in Eq.  \eqref{Pnks_2}, set $\kappa = 1$ and rewrite in the unordered coordinates as
\begin{align} \label{Pnk_app}
{\cal P}_{N,\kappa=1}(s) = \frac{1}{Z_N} \int_{\Gamma} \frac{d \tilde \mu}{2\pi i}  \int dx_1 \cdots dx_N \, \e^{\tilde \mu \left(s - \frac{1}{N}\sum_{i=1}^N x_i \right)} \e^{-\left[\frac{N^2}{2}\sum_{i=1}^N x_i^2 - \alpha \, N\,\sum_{i\neq i} |x_i-x_j| \right]} \;, \
\end{align}
where we recall that $Z_N$ is given by
\begin{align}\label{ZN_app}
Z_N = \int dx_1 \cdots dx_N \; \e^{-\left[\frac{N^2}{2}\sum_{i=1}^N x_i^2 - \alpha \, N\,\sum_{i\neq i} |x_i-x_j| \right]} \;.
\end{align}
In Eq. (\ref{Pnk_app}), we complete the squares for each $x_i$ and write it as
\begin{align} \label{Pnk_app2}
{\cal P}_{N,\kappa=1}(s) = \frac{1}{Z_N} \int_{\Gamma} \frac{d \tilde \mu}{2\pi i} \e^{\tilde \mu\,s}  \int dx_1 \cdots dx_N \, \e^{-\left[\frac{N^2}{2}\sum_{i=1}^N \left[\left(x_i -\frac{\tilde \mu}{N^3} \right)^2 - \frac{\tilde \mu^2}{N^6}\right]  - \alpha \, N\,\sum_{i\neq i} |x_i-x_j| \right]} \;.
\end{align}
By making a shift $x_i \to x_i - \tilde \mu/N^3$, and cancelling the partition function, we get
\begin{align} \label{Pnk_app3}
{\cal P}_{N,\kappa=1}(s) =  \int_{\Gamma}\,\frac{d \tilde \mu}{2\pi i} \, \e^{\tilde \mu\,s + \frac{\tilde \mu^2}{2N^3}} \;.
\end{align}
This is just a simple Gaussian integral, which can be easily evaluated, giving
\begin{align} \label{Pnk_app4}
{\cal P}_{N,\kappa=1}(s) = \frac{N^{3/2}}{\sqrt{2 \pi}}\, \e^{-\frac{N^3\, s^2}{2}} \;.
\end{align}
Thus for $\kappa= 1$, the probability distribution of the center of mass $s$ is a Gaussian with zero mean and variance $1/N^3$ and this Gaussian form actually
holds for all $s$.

\section{Typical fluctuations of the TLS}\label{sec:typical}

In this appendix, we show that the typical fluctuations of the TLS $s$ around its mean value $\langle s \rangle \simeq 2 \alpha \kappa(1-\kappa)$ is of order $O(N^{-3/2})$ and are described by a Gaussian distribution. 

We start from the exact representation of the distribution ${\cal P}_{N,\kappa}(s)$ in Eq. (\ref{Pnks_3}) in the main text, which reads
\begin{align} \label{Pnks_3_app}
{\cal P}_{N,\kappa}(s) = \frac{N^3\, N!}{Z_N} \int_{\Gamma} \frac{d \mu}{2\pi i} \,\e^{\mu N^3 s}\,  \int_< dx_1 \cdots dx_N \, \e^{-\beta E_\mu[\{x_i\}]} \;,
\end{align}
where the scaled energy $\beta E_\mu[\{x_i\}]$ is given in Eqs. (\ref{eq:mu_negative}) and (\ref{eq:const}). We set $s = \langle s \rangle + N^{-\varphi}\,z$
where $N^{-\varphi}$ denotes the scale of the typical fluctuations and the exponent $\varphi$ is yet to be determined. This gives, upon using the explicit form of the constant $C_1$ in (\ref{eq:const}) 
\begin{align} \label{Pnks_4_app}
{\cal P}_{N,\kappa}( \langle s \rangle + N^{-\varphi}\,z) \simeq  \frac{N^3\, N!}{Z_N} \e^{\frac{2 \alpha^2}{3}(N^2-N)}\,\int_{\Gamma} \frac{d \mu}{2\pi i} \, \e^{\mu N^{3-\varphi}\,z + N^3\,\mu^2\frac{\kappa}{2}}  \int_< dx_1 \cdots dx_N \e^{- {\cal F}_{\mu}[\{ x_i\}]}
\end{align}
where ${\cal F}_{\mu}[\{ x_i\}]$ is given in Eq. (\ref{calF_mu}). We now make a change of variable and set $\mu \, N^{3-\varphi}= p$. This gives
\begin{align} \label{Pnks_5_app}
{\cal P}_{N,\kappa}( \langle s \rangle + N^{-\varphi}\,z) \simeq  \frac{N^\varphi\, N!}{Z_N} \e^{\frac{2 \alpha^2}{3}(N^2-N)}\,\int_{\Gamma} \frac{d p}{2\pi i} \e^{p\,z + N^{2\varphi - 3}p^2 \frac{\kappa}{2}}  \int_< dx_1 \cdots dx_N \e^{- {\cal F}_{\mu=p N^{\varphi-3}}[\{ x_i\}]}.
\end{align}
Since we used $s = \langle s \rangle + N^{-\varphi}\,z$ for the typical fluctuations, we expect that ${\cal P}_{N,\kappa}( \langle s \rangle + N^{-\varphi}\,z)$ should scale like $N^\varphi \, f(z)$ where $f(z)$ is the scaling function that we are after. This means that the right hand side of (\ref{Pnks_5_app}), excluding the $N^\varphi$ factor, must be of order $O(1)$ as $N \to \infty$ -- in fact this should be just the scaling function $f(z)$. To get a nontrivial scaling function $f(z)$ of order $O(1)$, we must choose $\varphi = 3/2$ such that the term quadratic in $p$ inside the exponential is of order $O(1)$. Subsequently, one gets 
\begin{align} \label{Pnks_6_app}
{\cal P}_{N,\kappa}( \langle s \rangle + N^{-\frac{3}{2}}\,z) &\simeq  \frac{N^{\frac{3}{2}}\, N!}{Z_N} \,\int_{\Gamma} \frac{d p}{2\pi i} \e^{p\,z + p^2 \frac{\kappa}{2}}\int_< dx_1 \cdots dx_N \e^{- \left({\cal F}_{\mu=0}[\{ x_i\}]-\frac{2 \alpha^2}{3}(N^2-N)\right)}\;.
\end{align}
We now note that ${\cal F}_{\mu=0}[\{ x_i\}]-\frac{2 \alpha^2}{3}(N^2-N) = \beta E_{\mu=0}[\{x_i\}]$, using Eq. (\ref{eq:mu_negative}). But then the multiple integral is exactly $Z_N(\mu=0)/N!$, from Eq. (\ref{Zmu}). Using further $Z_N(\mu=0) = Z_N$ (see the discussion below Eq. (\ref{Zmu})), we get
\begin{align} \label{Pnks_7_app}
{\cal P}_{N,\kappa}( \langle s \rangle + N^{-\frac{3}{2}}\,z) &\simeq N^{\frac{3}{2}} \int_{\Gamma} \frac{d p}{2\pi i} \e^{p\,z + p^2 \frac{\kappa}{2}} \;.
\end{align}
Performing finally the Gaussian integral explicitly, we get 
\be \label{typ_Gaussian_app}
\mathcal{P}_{N, \kappa}(\langle s \rangle + N^{-\frac{3}{2}}\,z) \simeq N^{3/2} f_G\left( z \right)  \;,
\ee
where the scaling function $f_G(z)$ is given by
\be \label{Gaussian_form_app}
f_G(z) = \frac{1}{\sqrt{2 \pi \kappa}}\, \e^{- \frac{z^2}{2 \kappa}} \;.
\ee
This shows that the typical fluctuations of the TLS $s$, for any $0< \kappa \leq 1$, are of order $N^{-3/2}$ and are described by a scaling form (pure Gaussian),
as announced in Eqs. (\ref{typ_Gaussian}) and (\ref{Gaussian_form}) in the main text. 

Note that this derivation of the typical fluctuations holds for $\kappa > 0$, i.e., $\kappa = O(1)$ (independent of $N$), which selects the exponent $\varphi = 3/2$. If however $\kappa$ scales with $N$, as $\kappa = b/N$, with $b \geq 1$, then the argument of the exponential term in the $p$-integral in Eq. (\ref{Pnks_5_app}) reads: $p\,z + N^{2\varphi -4}p^2 b/2$. In this case, in order to make this term of order $O(1)$, we need to choose $\varphi  =2$. In that case, it turns out that the multiple integral over $x_i$'s in Eq.~(\ref{Pnks_5_app}) has a limiting value, independent of $N$. In that case, the Laplace transform of the ${\cal P}_{N, \kappa}(s)$ has a non-trivial scaling form 
\begin{align} \label{scaling_b}
{\cal P}_{N, \kappa = b/N}(s) \simeq N^2 f_{\alpha, b}\left( (s-\langle s \rangle)\, N^2\right) \;,
\end{align}
where the scaling function $f_{\alpha, b}(z)$ is parametrised by $\alpha$ and $b \geq 1$. In particular, in the limit $b \to 1$, which corresponds to $s = x_{\max}/N$, we have verified that 
the scaling function $f_{\alpha,1}(z)$ coincides with the known limiting distribution of $x_{\max}/N$, computed in Refs. \cite{Dhar2017, Dhar2018}.

\section{Computation of an ordered sum}\label{sec:appsum}

In this appendix, we want to establish the identity 
\bea \label{id}
\sum_{i>j}(x_i-x_j) = \sum_{i=1}^N(2i-N-1) \,x_i \;.
\eea
We consider the first term on the left hand side of Eq. (\ref{id}) and write it as
\bea \label{T1}
T_1 = \sum_{i>j} x_i = \sum_{i=1}^N x_i \sum_{j=1}^{i-1} 1 = \sum_{i=1}^N (i-1)\, x_i \;.
\eea
Now we consider the second term on the left hand side of (\ref{id})
\bea \label{T2}
T_2 = \sum_{i>j} x_j = \sum_{j=1}^N x_j \sum_{i=j+1}^N 1 = \sum_{j=1}^N (N-j) \,x_j = \sum_{i=1}^N (N-i) \,x_i \;.
\eea
Subtracting $T_2$ from $T_1$ gives Eq. (\ref{id}).

\end{document}